\RequirePackage{fix-cm}
\documentclass[twocolumn]{svjour3}          
\smartqed  
\usepackage{graphicx}
\usepackage{amssymb}
\begin{document}

\title{Gate voltage tuned quantum superconductor to insulator transition in
an ultrathin bismuth film revisited}

\author{T.~Schneider \and S.~Weyeneth}

\institute{T.~Schneider \at Physik-Institut der Universit\"at Z\"urich, Winterthurer\-strasse~190, CH-8057 Z\"urich, Switzerland \\\email{toni.schneider@swissonline.ch} 
\and
S.~Weyeneth \at Physik-Institut der Universit\"at Z\"urich, Winterthurer\-strasse~190, CH-8057 Z\"urich, Switzerland \\}

\date{Received: date / Accepted: date}

\maketitle

\begin{abstract}
We explore the implications of Berezinskii-Kosterlitz-Thouless (BKT) critical behavior and variable-range hopping on the two dimensional (2D) quantum superconductor-insulator (QSI) transition driven by tuning the gate voltage. To illustrate the potential and the implications of this scenario we analyze sheet resistance data of Parendo \textit{et al.} taken on a gate voltage tuned ultrathin amorphous bismuth film. The finite size scaling analysis of the BKT-transition uncovers a limiting length $L$ preventing the correlation length to diverge and to enter the critical regime deeply. Nevertheless the attained BKT critical regime reveals consistency with two parameter quantum scaling and an explicit quantum scaling function determined by the BKT correlation length. The two parameter scaling yields for the zero temperature critical exponents of the QSI-transition the estimates $z\overline{\nu }\simeq 3/2$, $z\simeq 3$ and $\overline{\nu } \simeq 1/2$, revealing that hyperscaling is violated and in contrast to finite temperature disorder is relevant at zero temperature. Furthermore, $z\overline{\nu }\simeq 3/2$ is also consistent with the two variable quantum scaling form associated with a variable-range hopping controlled insulating ground state.

\keywords{Superconducting films \and Quantum phase transition \and Two parameter scaling \and Superconductor insulator transition}
\PACS{74.40.Kb \and 71.30.+h \and 72.15.Rn \and 74.78.-w}
\end{abstract}

\section{Introduction}
\label{sec:1}

Continuous quantum-phase transitions (QPT) are transitions at absolute zero in which the ground state of a system is changed by varying a parameter of the Hamiltonian \cite{sondhi,tsjs,tsben}. The transitions between superconducting and insulating behavior in two-dimensional systems tuned by disorder, film thickness, magnetic field or with the electrostatic field effect are believed to be such transitions \cite{tsjs,markovic,gant}.

Here we present a detailed analysis of the temperature and gate voltage dependent sheet resistance data of Parendo \textit{et al.} \cite{parendo,parendo2}. to explore in an ultrathin amorphous bismuth film, the nature of the normal state to superconductor (NS) phase transition line including its end point at zero temperature, as well as the normal state to insulator (NI) crossover. At the common endpoint of the two lines the film is supposed to undergo a quantum superconductor to insulator (QSI) transition, separating the superconducting from the insulating ground state.

Considering the NS- transition line we explore the compatibility with gate voltage dependent Berezinskii-Kosterlitz-Thouless (BKT) critical behavior \cite{bere,kosterlitz}. Our analysis of the temperature dependence of the sheet resistance at various fixed gate voltages uncovers a round\-ed BKT-transition. The rounding is fully consistent with a standard finite-size effect whereupon the correlation length is prevented to grow beyond a limiting length $L$. Potential candidates for the limiting length include the failure in cooling \cite{parendo,parendo2}, limited homogeneity due to local strain or a heat current. A nonzero heat current drives the system away from equilibrium and creates a temperature gradient which implies a space dependent temperature. Because the correlation length does not exhibit the usual and relatively slow algebraic divergence as $T_{c}$ is approached, the BKT-transition is particularly susceptible to such finite-size effects. On the other hand there is the Harris criterion \cite{harris,aharony}, stating that short-range correlated and uncorrelated disorder is irrelevant at the unperturbed critical point, provided that $\nu >2/D$, where $D$ is the dimensionality of the system and $\nu$ the critical exponent of the finite-temperature correlation length. With $D=2$ and $\nu =\infty$, appropriate for the BKT-transition \cite{bere,kosterlitz}, this disorder should be irrelevant. In this context it is important to recognize that the existence of the BKT-transition (vortex-antivortex dissociation instability) in $^{4}$He films is intimately connected with the fact that the interaction energy between vortex pairs depends logarithmic on the separation between them. As shown by Pearl \cite{pearl} vortex pairs in thin superconducting films (charged superfluid) have a logarithmic interaction energy out to the characteristic length $\lambda _{2D}=\lambda ^{2}/d$, beyond which the interaction energy falls off as $1/r$. Here $\lambda$ is the magnetic penetration depth of the bulk. As $\lambda _{2D}$ increases the diamagnetism of the superconductor becomes less important and the vortices in a thin superconducting film become progressively like those in $^{4}$He films \cite{beasley}. According to this $\lambda _{2D}>>\min \left[ W,~L\right] $ is required, where $W$ and $L$ denote the width and the length of the perfect sample. Invoking the Nelson-Kostelitz relation \cite{nelson} $\lambda_{2D}\left( T_{c}\right) =\lambda ^{2}\left( T_{c}\right) /d=\Phi_{0}^{2}/\left( 32\pi ^{2}k_{B}T_{c}\right)$ it is readily seen that for sufficiently low $T_{c}$'s and $\min \left[ W,~L\right] <<1$ cm this condition is well satisfied. As a result a rounded transition uncovers a limiting length which is shorter than that resulting from the finite magnetic "screening length" $\lambda _{2D}$. Nevertheless, for sufficiently large $L$ the critical regime can be attained and a finite-size scaling analysis provides good approximations for the limit of fundamental interest, $L\rightarrow \infty$ \cite{privman}.

As will be shown below, the finite-size scaling analysis uncovers rounded NS- phase transitions along the line $T_{c}(V_{g})$, ending and vanishing at the critical gate voltage $V_{gc}$ where the QSI transition occurs. The critical properties along this line are fully consistent with BKT-behavior subjected to a gate voltage dependent limiting length which decreases substantially by approaching the QSI critical point at $V_{gc}$. According to this and in agreement with previous studies \cite{tsprb,tssw}, electrostatic tuning does not change the carrier density only but the inhomogeneity landscape as well. Furthermore it is shown that the BKT-behavior leads to an explicit and extended expression of the standard quantum scaling form of the sheet resistance \cite{sondhi,tsjs,tsben}
\begin{equation}
\frac{R\left( T,V_{g}\right)}{R_{c}}=G_{\pm }\left( y\right) ,y=\frac{c\left\vert V_{g}-V_{gc}\right\vert ^{z\overline{\nu }}}{T}.  
\label{eq1a}
\end{equation}
The subscripts $\pm$ denote the branch resulting from the the NI-crossover (+) and NS-transition (-), respectively $c$ is a nonuniversal coefficient of proportionality, the gate voltage $V_{g}$ the tuning parameter with critical value $V_{gc}$. $z$ is the dynamic and $\overline{\nu }$ the critical exponent of the zero temperature correlation length, $\xi \left(T=0,V_{g}\right) =\xi _{0}\left( T=0\right) \left\vert \delta -\delta_{c}\right\vert ^{-\overline{\nu }}$. $G_{\pm}\left(y\right)$ is a universal scaling function of its argument such that $G_{\pm }\left(y=0\right) =1$. The BKT-behavior yields with the variables $R_{0-}\left(V_{g}\right) $ and $y$ the two parameter scaling form 
\begin{equation}
\frac{R\left( T,V_{g}\right) }{R_{0\pm }\left( V_{g}\right) }=G_{\pm }\left(y\right). 
\label{eq1b}
\end{equation}
Noting that $R_{0-}\left( V_{g}\right) $ exhibits a substantial gate voltage dependence the one parameter scaling form (\ref{eq1a}) applies asymptotically only ($R_{0-}\left( V_{g}\right) \rightarrow R_{c}$). As the film undergoes at finite temperature a BKT-transition the scaling function $G_{-}\left( y\right)$ exhibits at the universal value $y_{c}$ a finite temperature singularity. The leading behavior of the phase transition line is then fixed by
\begin{equation}
T_{c}\left( V_{g}\right) =\frac{c}{y_{c}}\left\vert V_{g}-V_{gc}\right\vert^{z\overline{\nu }}.  
\label{eq1c}
\end{equation}
Contrariwise the BKT expression for the sheet resistance yields the explicit scaling function
\begin{equation}
G_{-}\left( y\right) =\exp \left( -\widetilde{b}_{R}\left( T/T_{c}\left(V_{g}\right) -1\right) ^{-1/2}\right),  
\label{eq1e}
\end{equation}
which applies for any $T\geq T_{c}$. $\widetilde{b}_{R}$ is a non-universal dimensionless constant. In this context it is essential to notice that the universal critical behavior is entirely classical for $T_{c}>0$ \cite{hertz}, including the characteristic form of the BKT correlation length which determines $G_{-}\left( y\right)$. Quantum fluctuations enter via renormalization of $R_{0-}\left( V_{g}\right) $, the constraint $R_{0-}\left( V_{g}\right) \rightarrow R_{0-}\left( V_{gc}\right) =R_{c}$ and $T_{c}\left( V_{g}\right) $. In contrast to the standard scaling form (\ref{eq1a}) it applies for any $T\geq T_{c}\left( V_{g}\right) $ down to the critical endpoint of the BKT-line where Eq. (\ref{eq1c}) holds because $T/T_{c}$ approaches $T/T_{c}=y_{c}/y$. Using the finite size corrected estimates for $R_{0-}\left( V_{g}\right) $ and the BKT-line $T_{c}\left(V_{g}\right) $ we observe in an intermediate range of the scaling variable $T/T_{c}\left( V_{g}\right) $ a satisfactory collapse of the sheet resistance data onto the BKT scaling function (\ref{eq1e}). The lower bound of this range stems from temperatures considerably above $T_{c}$ where BKT fluctuations no longer dominate and the upper bound is traced back to a finite size effect preventing the correlation length to grow beyond a limiting length $L\left( V_{g}\right)$. Taking the substantial gate voltage dependence of $R_{0-}$ into account we observe close to the QSI-transition consistency with $T_{c}\left( V_{g}\right) \propto \left\vert V_{g}-V_{gc}\right\vert ^{z\overline{\nu }}$ (Eq. (\ref{eq1c})) for $z\overline{\nu }\simeq 3/2$ and the corresponding expression for the BKT-scaling function
\begin{equation}
G_{-}\left( y\right) =\exp \left( -\widetilde{b}_{R}\left( y_{c}/y-1\right)^{1/2}\right) .  
\label{eq1f}
\end{equation}
The estimate $z\overline{\nu }\simeq 3/2$ coincides with the value obtained from the analysis of the sheet resistance data of gate voltage tuned epitaxial films of La$_{2-x}$Sr$_{x}$CuO$_{4}$ that are one unit cell thick \cite{tssw,bol}.

To estimate the exponents $z$ and $\overline{\nu }$ we note that according to the finite temperature dynamic scaling theory $R_{0-}\left( V_{g}\right)$ is proportional to $\xi _{0}^{-2}\left( V_{g}\right) $ where $\xi _{0}\left(V_{g}\right) $ is the amplitude of the BKT correlation length 
\begin{eqnarray}
& {\rm } &\xi \left( T,V_{g}\right)= \nonumber\\
& {\rm } &\xi _{0}\left( V_{g}\right) \cdot  \exp \left( -\left(\widetilde{b}_{R}/2\right) \left( T/T_{c}\left( V_{g}\right) -1\right)^{-1/2}\right).
\label{eq1g}
\end{eqnarray}
Neglecting the quantum constraint of a finite critical sheet resistance $R_{c}$ the amplitude $R_{0-}\left( V_{g}\right) \propto \xi _{0}^{-2}\left(V_{g}\right) $ should match the gate voltage dependence of the zero temperature counterpart $\xi ^{-2}\left( T=0,V_{g}\right)$ which tends to zero at the critical gate voltage because $\xi \left( T=0,V_{g}\right)$ diverges as $\xi \left( T=0,V_{g}\right) =\xi _{0}\left( T=0\right) \left(V_{g}-V_{gc}\right) ^{-\overline{\nu }}$. The quantum constraint is then accounted for in terms of \cite{tsprb,tssw}
\begin{eqnarray}
R_{0-}(V_{g})-R_{c} &\propto& \xi _{0}^{-2}\left( V_{g}\right) \propto \xi^{-2}\left( T=0,V_{g}\right)  \nonumber\\
&\propto&\left( V_{g}-V_{gc}\right) ^{-2\overline{\nu }}\propto T_{c}^{2/z}.  
\label{eq1h}
\end{eqnarray}
Using the estimates for $R_{0-}(V_{g})$, $R_{c}$ and $T_{c}\left(V_{g}\right)$ resulting from the finite size scaling analysis of the temperature dependence of the sheet resistance at fixed gate voltages this relationship provides a direct way to measure the critical exponents $z$ and $\overline{\nu }$, and allows for a check on the value of $z\overline{\nu }$. We obtain $z\simeq 3$ and $\overline{\nu }\simeq 1/2$. Noting that $D+z=2+z\simeq 5$ exceeds the upper critical dimension $D_{u}=4$ the critical exponent of the zero temperature correlation length should indeed adopt its mean-field value $\overline{\nu }=1/2$. However the fate of this clean QSI critical point under the influence of disorder is controlled by the Harris criterion \cite{harris,aharony}. If the inequality $\overline{\nu }\geq 2/D$ is fulfilled, the disorder does not affect the critical behavior. If the Harris criterion is violated ($\overline{\nu }<2/D$), the generic result is a new critical point with conventional power law scaling but new exponents which fulfill $\overline{\nu }<2/D$ and $\overline{\nu }\neq 1/2$. Because disorder is relevant it drives the system from the mean-field to an other critical point with different critical exponents and with that to $\overline{\nu }\neq 1/2$. Unfortunately the available data are too sparse to derive a more precise estimate of $\overline{\nu }$. Otherwise it is clear that $D+z=2+z\simeq 5$ exceeds the upper critical dimension $D_{u}=4$ so that the the equivalence between quantum phase transitions in systems with $D$ spatial dimensions and the ones of classical phase transitions in $(D+z)$ dimensions does not apply.

Concerning the NI-crossover it is known that in Bi-films spin-orbit coupling is important because its strength strongly depends on the atomic number Z and bismuth is the heaviest group-V semimetal \cite{marcano}. In the presence of strong spin-orbit coupling, the spins rotate in the opposite direction and the interference is destructive, which leads a reduction of weak localization, known as weak anti-localization \cite{bergmann}. Given the previous evidence for this behavior in Bi-films \cite{marcano} one might expect that weak anti-localization controls the NI-crossover. Otherwise it is also expected that in disordered films variable-range hopping controls the insulating phase in terms of $\sigma \left( T,V_{g}\right) =\sigma_{h}\left( V_{g}\right) \exp \left( -\left( T_{0}\left( V_{g}\right)/T\right) ^{1/3}\right)$ \cite{mott}, transforming to the scaling function 
\begin{equation}
G_{+}\left( y\right) \simeq \exp (fy^{1/3}),  
\label{eq1i}
\end{equation}
with the crossover temperature $T_{0}\left( V_{g}\right) \propto f^{3}\left\vert V_{g}-V_{gc}\right\vert ^{z\overline{\nu }}$. In this context it should be kept in mind that the irrelevance of disorder at the BKT-transition does not exclude its relevance at zero temperature. Indeed our estimate $\overline{\nu }\simeq 1/2$ violates the Harris criterion \cite{harris,aharony} and with that is the film considered here at $T=0$ sensitive to disorder. For this reason we assume that the insulating ground state is controlled by variable range hopping. We observe that the associated quantum scaling function (Eq. (\ref{eq1i})), combined with a strongly gate voltage dependent $\sigma _{h}\left( V_{g}\right)=1/R_{0+}\left( V_{g}\right) $ mimics the sheet resistance data remarkably well, suggesting that the insulating phase of the Bi film considered here is controlled by variable-range hopping.

Essential conclusions of our analysis include: As \linebreak $R_{0-}(V_{g})=R_{0}(V_{g})$ and $R_{0+}(V_{g})$ exhibit a strong gate voltage dependence is two parameter scaling with the variables $R_{0\pm }\left( V_{g}\right) $ and $y=c\left\vert V_{g}-V_{gc}\right\vert ^{z\overline{\nu }}/T$ unavoidable to attain the QSI-transition. The piecewise data collapse onto the respective scaling function $G_{\pm }$ $\left( y\right) $ extends over two decades of the scaling variable. The BKT-line leads to an explicit expression of the scaling function $G_{-}$ $\left( y\right) $ and the evidence for a variable-range hopping controlled insulating phase determines $G_{+}\left(y\right) $. For large values of the scaling argument the deviations from these scaling functions are traced back to a finite size effect with a limiting length which increases with reduced gate voltage. In the lower branch $G_{-}$ $\left( y\right) $ deviations occur for small arguments as well because BKT fluctuations no longer dominate. The scaling function of the upper and lower branch are very different but both exhibit at the QSI-transition an essential singularity. Our estimates for the critical exponents at the QSI critical point are consistent with the following properties: $z\overline{\nu }\simeq 3/2$ emerging from the BKT-line yields to a satisfactory data collapse onto both branches of the quantum scaling
function $G_{\pm }$ $\left( y\right) $. The estimates $z\simeq 3$ and $\overline{\nu }\simeq 1/2$ emerge from the gate voltage dependence of $R_{0-}(V_{g})-R_{c}$. The Harris criterion \cite{harris,aharony} implies however that $\overline{\nu }=1/2$ is excluded, disorder is relevant at zero temperature but irrelevant an finite temperature. Furthermore it is shown that the application of one parameter scaling form(\ref{eq1a}) leads to rather a different $z\overline{\nu }$ value because the regime where $R_{0-}(V_{g})\simeq R_{c}$ applies is not attained,  because $R_{0-}(V_{g})$ is a nonuniversal function entering the BKT-scaling form of the sheet resistance. Nevertheless its value at the QSI- transition, $R_{0-}(V_{gc})=R_{c}$ is expected to be universal and given $R_{c}=c_{R}h/4e^{2}\simeq c_{R}\cdot6.45$ k${\rm \Omega}$ \cite{fisher,weich,fishweich}. $c_{R}$ is a dimensionless constant. Provided that hyperscaling applies it depends only on the universality class of the QSI transition \cite{weich}. In the present case does hyperscaling not apply, because $D+z=2+z\simeq 5$ exceeds the upper critical dimension $D_{u}=4$. In any case our estimate $R_{c}\simeq 19.35$ k${\rm \Omega} $ is of order $h/e^{2}\simeq 25.8$ k${\rm \Omega}$.

The nature of the two-dimensional QSI-transition has been intensely debated \cite{markovic,gant,weich}. The scenarios can be grouped into two classes, fermionic and bosonic. In the fermionic case the reduction of $T_{c}$ and the magnitude of the order parameter is attributed to a combination of reduced density of states, enhanced Coulomb interaction and depairing due to an increase of the inelastic electron-electron scattering rate \cite{maekawa,finkel}. The bosonic approach assumes that the fermionic degrees of freedom can be integrated out, the mean square of the order parameter does not vanish at $T_{c}$, phase fluctuations dominate and the reduction of $T_{c}$ is attributable to quantum fluctuations and in disordered systems to randomness in addition \cite{weich,fishweich,fishgrin}. In the Bi-film considered here the consistency with finite size limited BKT behavior uncovers clearly the bosonic scenario: The BKT-line separates the phases of uncondensed and condensed Cooper pairs, while at zero temperature the QSI critical point separates the phase of condensed and uncondensed insulating Cooper pairs.

In Sec. II we sketch the theoretical background and present the detailed analysis of the resistivity data of Parendo \textit{et al.} \cite{parendo,parendo2}. We close with a brief summary and some discussion.

\section{Theoretical background and data analysis}
\label{sec:2}

\subsection{BKT transition}
\label{sec:2:1}

To explore the compatibility with BKT critical behavior we invoke the characteristic temperature dependence of the correlation length above $T_{c}$ \cite{bere,kosterlitz}, 
\begin{equation}
\xi \left( T,V_{g}\right) =\xi _{0}\left( V_{g}\right) \exp \left( \frac{b_{R}\left( V_{g}\right) }{2\left( T-T_{c}\left( V_{g}\right) \right) ^{1/2}}\right), 
\label{eq1}
\end{equation}
Notice that for any $T_{c}>0$ the universal critical behavior is entirely classical \cite{hertz}, including the characteristic form of the BKT correlation length. Contrariwise $\xi _{0}\left( \delta \right) $, $b_{R}\left( \delta \right) $ and $T_{c}\left( \delta \right) $ are subjected to quantum fluctuation renormalization and depend on the tuning parameter $\delta $. $b_{R}$ is related to the vortex core energy and $\xi _{0}$ to the vortex core radius \cite{tsprb,dahm,steel,williams}. Invoking dynamic scaling the sheet resistance $R$ scales in $D=2$ as \cite{tsjs} 
\begin{equation}
R\propto \xi ^{-z_{cl}},  
\label{eq2}
\end{equation}
where $z_{cl}$ is the dynamic critical exponent of the classical dynamics. $z_{cl}$ is not questioned to be anything but the value that describes simple diffusion: $z_{cl}=2$ \cite{pierson,tstool}. Combining these scaling forms we obtain
\begin{eqnarray}
\frac{R\left( T,\delta \right) }{R_{0-}\left( \delta \right) }&=&\left( \frac{\xi _{0}\left( V_{g}\right) }{\xi \left( T,V_{g}\right) }\right) ^{2} \nonumber\\
&=&\exp\left( -\frac{b_{R}\left( V_{g}\right) }{\left( T-T_{c}\left( V_{g}\right)\right) ^{1/2}}\right) .
\label{eq3}
\end{eqnarray}
The compatibility of sheet resistance data with the characteristic BKT behavior can be explored in terms of
\begin{eqnarray}
\left(\frac{d\ln \left( R\left( T,V_{g}\right) \right) }{dT}\right)^{-2/3}&=&\nonumber\\
\left( \frac{2}{b_{R}\left( V_{g}\right) }\right) ^{2/3} &\cdot& \left(T-T_{c}\left( V_{g}\right) \right). \label{eq4}
\end{eqnarray}
However, supposing that there is a limiting length $L$ preventing the growth of the correlation length the transition temperature $T_{c}$ cannot be approached and with that the attainable critical regime is reduced. The extent of the attained BKT-regime can be extracted from a finite size scaling analysis. It implies that the measured $R\left( T,L\right) $ scales as \cite{privman} 
\begin{equation}
\frac{R\left( T,L\right) }{R\left( T,L=\infty \right) }=\left( \frac{\xi\left( T,L=\infty \right) }{\xi \left( T,L\right) }\right) ^{2}=f(x),
\label{eq5}
\end{equation}
where
\begin{equation}
x=R^{-1}\left( T,L=\infty \right) \propto \xi ^{-2}\left( T,L=\infty \right) 
\label{eq6}
\end{equation}
$R\left( T,L=\infty \right) $ and $\xi \left( T,L=\infty \right) $ denote the respective behavior of the infinite and homogeneous system given by Eqs. (\ref{eq1}) and (\ref{eq3}), respectively. $f(x)$ is the finite size scaling function adopting the limiting behavior:
\begin{equation}
f(x)=\left\{
\begin{array}{c}
1~:\xi \left( T,\infty \right) <<L \\ 
g\left( L\right) x:\xi \left( T,\infty \right) >>L
\end{array}
\right.   
\label{eq7}
\end{equation}
Accordingly, BKT-behavior, $R\left( T,L\right) =R\left( T,L=\infty \right)$ can be observed as long as $\xi \left( T,\infty \right) <L$ , while for $\xi\left( T,\infty \right) >L$ the scaling function approaches the finite size dominated regime where 
\begin{equation}
R\left( T,L\right) \propto g\left( L\right) \propto L^{-2}.  
\label{eq8}
\end{equation}
Note that $g\left( L\right) $ fixes the width of the intermediate re\-gime where BKT-behavior is observable. It shrinks with increasing $g\left(L\right) $, \textit{i.e.} with decreasing limiting length $L$. This power law neglects the possible occurrence of multiplicative logarithmic correction \textit{i.e.} $R\left( L\right) \propto g\left( L\right) \propto L^{-2}/\ln \left( L/b_{0}\right) $.\cite{medv} Noting that the finite size scaling plot yields $g\left( L\right) $, this correction leads to a saturation of the sheet resistance at low temperatures as well and that the data considered here is too sparse to uncover it, we neglect this correction. The occurrence of BKT-behavior also requires that $T_{c}$ is
sufficiently below the mean-field transition temperature $T_{c0}$. It can be estimated with the aid of the Aslamosov-Larkin (AL) expression for the conductance \cite{aslamosov}
\begin{equation}
\sigma =\sigma _{n}+\sigma _{0}/\ln \left( T/T_{c0}\right) ,  
\label{eq9}
\end{equation}
with
\begin{equation}
\sigma _{0}=\frac{\pi e^{2}}{8h}\simeq 0.0152~{\rm k}{\rm \Omega} ^{-1},
\label{eq10}
\end{equation}
where Gaussian fluctuations are taken into account.

\begin{figure}
\includegraphics[width=\linewidth]{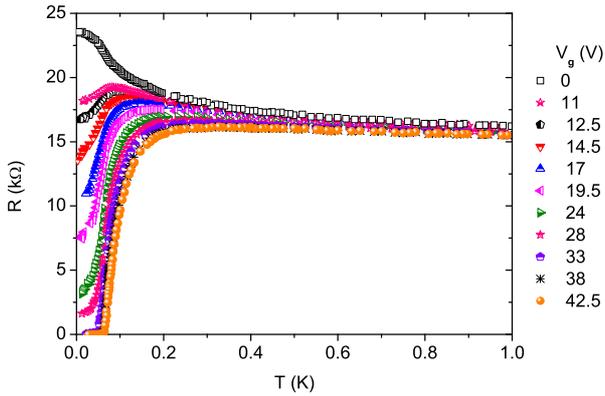}
\caption{$R$ \textit{vs}. $T$ for various gate voltages $V_{g}$ for the $10.22$ \AA\ film taken from Parendo \textit{et al.} \cite{parendo,parendo2} for various gate voltages $V_{g}$.}
\label{fig1}       
\end{figure}

\begin{figure}
\includegraphics[width=\linewidth]{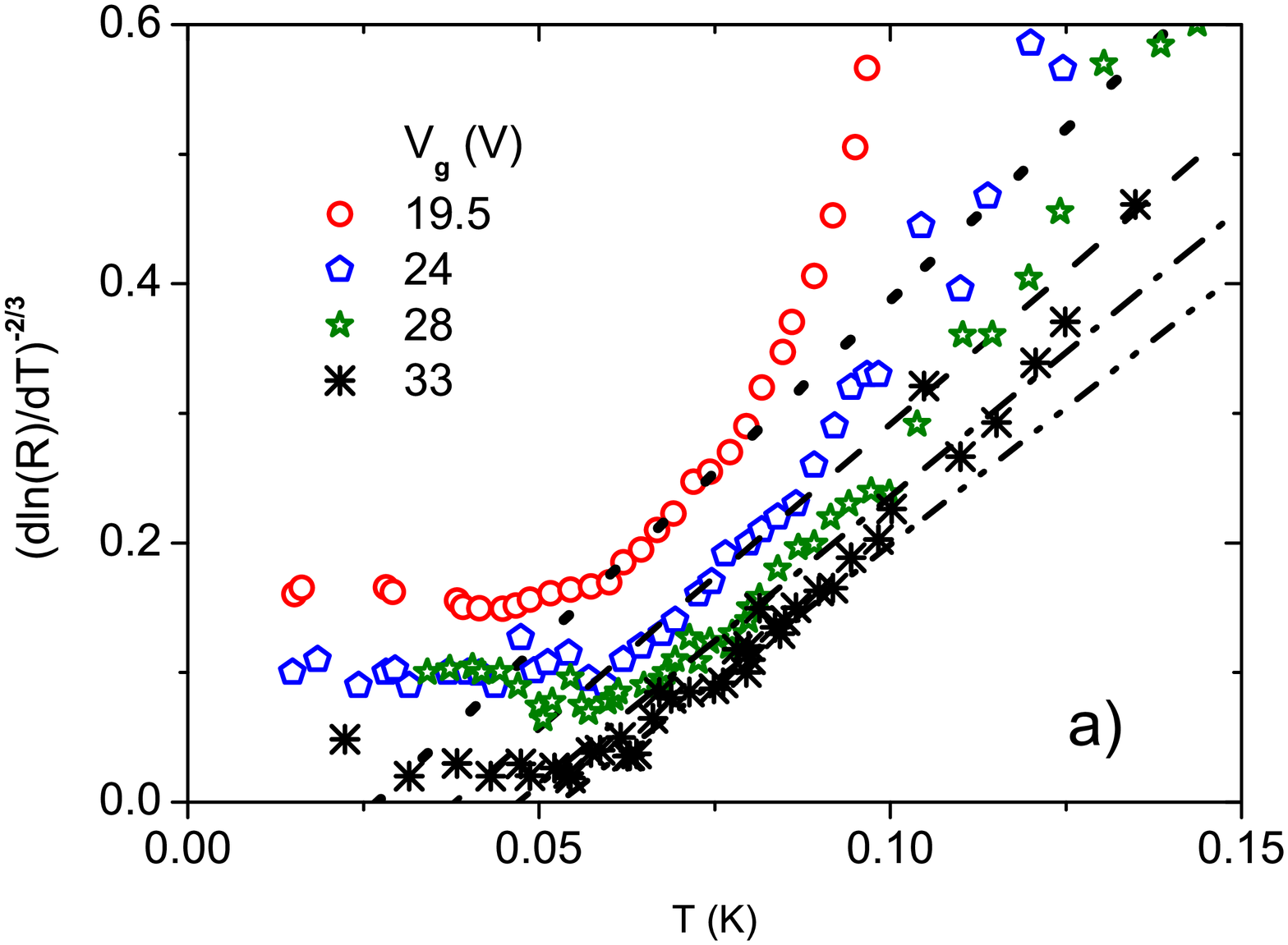}
\includegraphics[width=\linewidth]{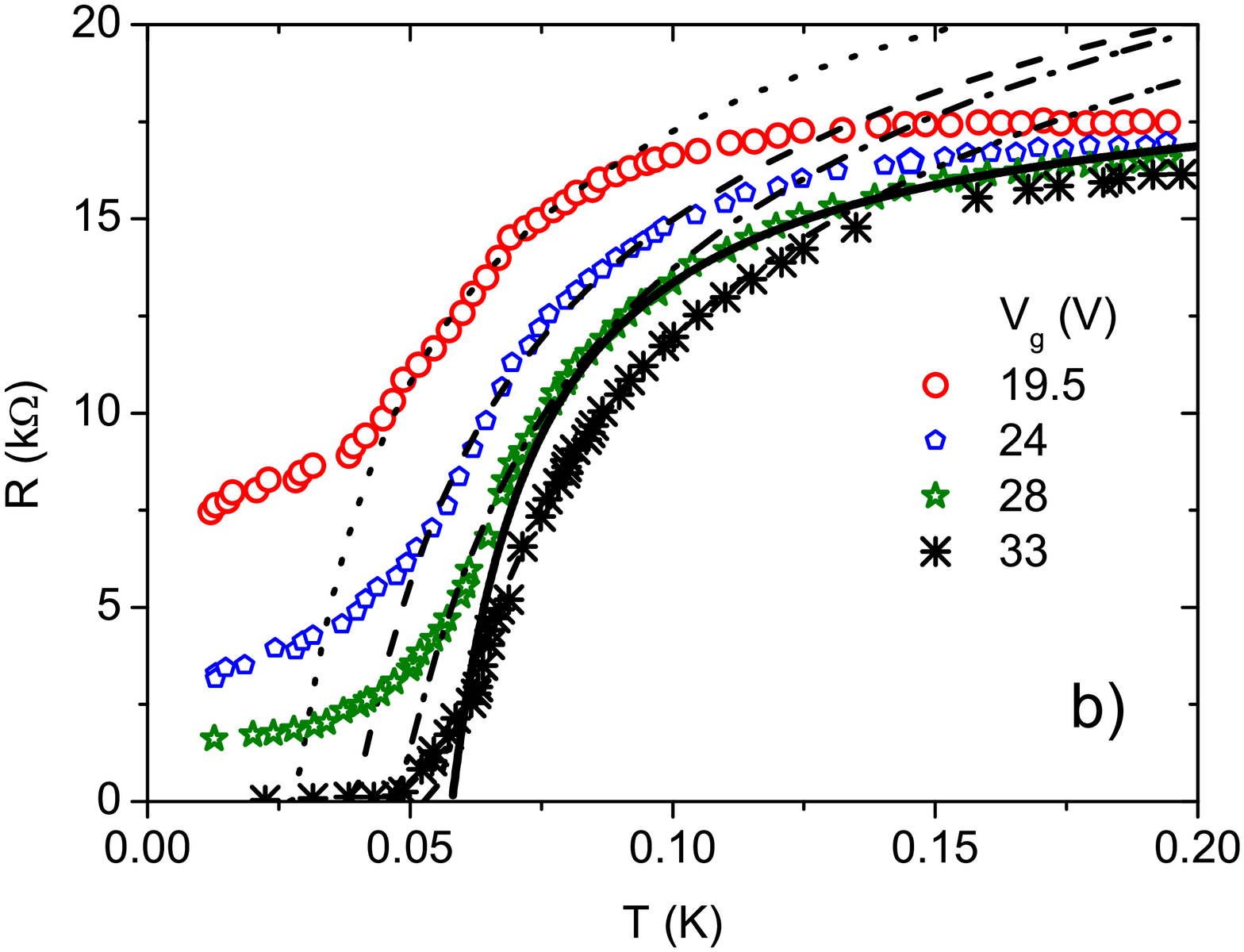}
\includegraphics[width=\linewidth]{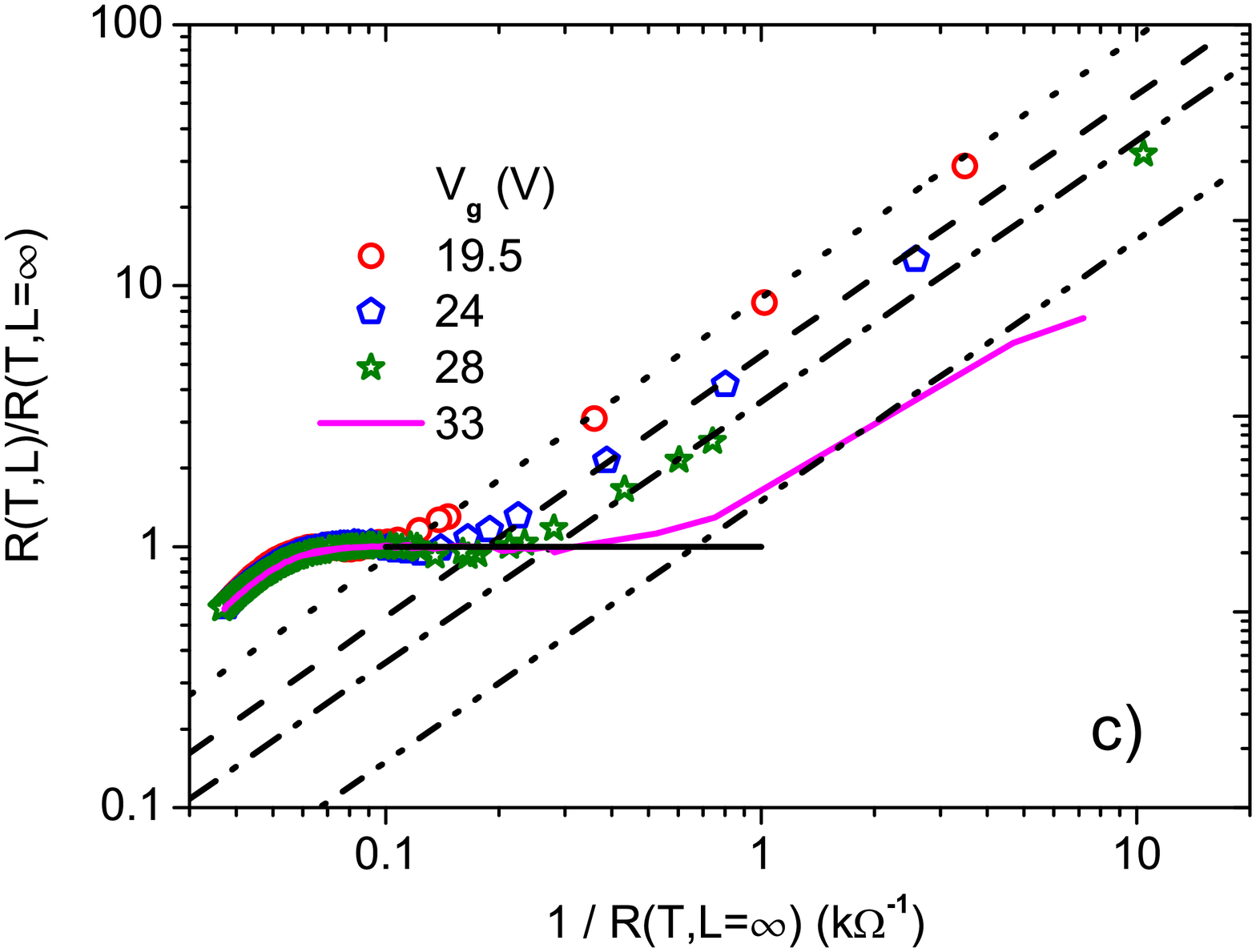}
\caption{a) $\left(d{\rm ln}(R)/dT\right) ^{-2/3}$ \textit{vs}. $T$ derived from the data of Parendo \textit{et al.} \cite{parendo,parendo2}. The lines are fits to Eq. (\ref{eq4}) yielding estimates for $T_{c}$ and $\left(2/b_{R}\right)^{2/3}$. ($T_{c}=0.027$~K and $\left( 2/b_{R}\right)^{2/3}=5.3$~K$^{-1/3}$.) b) $R$ \textit{vs}. $T$ for the same gate voltages. The lines are Eq. (\ref{eq1}) with $T_{c}$, $b_{R}$ derived from Fig. 2a and adjusted $R_{0}$ to match the respective sheet reistance data above $T_{c}$. The solid line is a fit to the Aslamosov-Larkin expression (\ref{eq9}) yielding $\sigma _{n}=0.044$~k${\rm \Omega} ^{-1}$ and $T_{c0}=0.04$~K compared to the corresponding BKT transition temperature $T_{c}=0.027$~K. c) Finite size scaling plot $R\left(T,L)/R(T,L=\infty \right) $ \textit{vs}. $1/R(T,L=\infty )$. The solid line marks the occurrence of BKT-behavior and the dash-dot line the finite size dominated regime where $R\left( T,L\right) \propto g\left( L\right) \propto L^{-2}$.}
\label{fig2}       
\end{figure}

We are now prepared to explore the occurrence of BKT-behavior emerging from the sheet resistance data of Parendo \textit{et al.} \cite{parendo,parendo2} depicted in Fig.~\ref{fig1} for an amorphous $10.22$~\AA\ thick Bi film taken at various gate voltages. It is apparent that most sheet resistance curves exhibit a rounded normal state to superconductor transition. Nevertheless a reduction of the smeared transition temperature with decreasing gate voltage can be anticipated. Accordingly BKT-behavior is expected in an intermediate temperature regime above $T_{c}$ only. and with that . To uncover this regime we proceed as follows: Invoking Eq. (\ref{eq4}) consistency with BKT-behavior is established in terms of an intermediate temperature regime revealing a linear temperature dependence for the adjusted parameters $\left(2/b_{R}\left( V_{g}\right) \right) ^{2/3}$ and $T_{c}\left( V_{g}\right) $. The amplitude $R_{0}\left( V_{g}\right) $ is then estimated by adjusting Eq. (\ref{eq3}) in the intermediate temperature regime to the measured temperature dependence of the sheet resistance with given $\left(2/b_{R}\left( V_{g}\right) \right) ^{2/3}$ and $T_{c}\left( V_{g}\right) $. The quality of the resulting estimates for $\left( 2/b_{R}\left(V_{g}\right) \right) ^{2/3}$, $T_{c}\left( V_{g}\right) $ and $R_{0}\left(V_{g}\right) $ as well as the evidence for finite size limited BKT behavior is finally clarified in terms of the finite size scaling relation (\ref{eq5}). Results of this analysis are shown in Fig.~\ref{fig2}a uncovers consistency with BKT-behavior (\ref{eq4}), $i.e.$ for $V_{g}=19.5$~V with $\left( 2/b_{R}\right) ^{2/3}\simeq 5.3$~K$^{-1/3}$ and $T_{c}\simeq 0.027$ K in the intermediate regime extending from $T\simeq 0.055$~K to $T=0.08$~K. Above this regime BKT-fluctuations no longer dominate, while below its lower bound a pronounced finite size effect occurs. These features are well confirmed in Fig.~\ref{fig2}b showing $R$ \textit{vs}. $T$. Indeed, the curves corresponding to Eq. (\ref{eq3}) fit the data in the intermediate temperature regime with the adjsuted $R_{0}=33.5$~k${\rm \Omega} $, $\left(2/b_{R}\right) ^{2/3}\simeq5.3$~K$^{-1/3}$ and $T_{c}\simeq 0.027$~K rather well. Furthermore the BKT-transition temperature is considerably below the mean-field counterpart as illustrated for $V_{g}=19.5$~V where $T_{c}\simeq0.027$~K compared to $T_{c0}\simeq0.04$~K. Considering the finite size scaling plot in Fig.~\ref{fig2}c it becomes clear that the rounding of the transition is fully consistent with a standard finite size effect, preventing the correlation length to grow beyond a limiting length $L$. Noting that the intermediate regime where BKT-behavior is observable shrinks with increasing $g$, the rather large value $g=9$~k${\rm \Omega}$ at $V_{g}=19.5$~V is consistent with the severely rounded transition. Nevertheless, in agreement with Fig.~\ref{fig2}a and Fig.~\ref{fig2}b there is a window left where BKT-behavior occurs. 

\begin{figure}
\includegraphics[width=\linewidth]{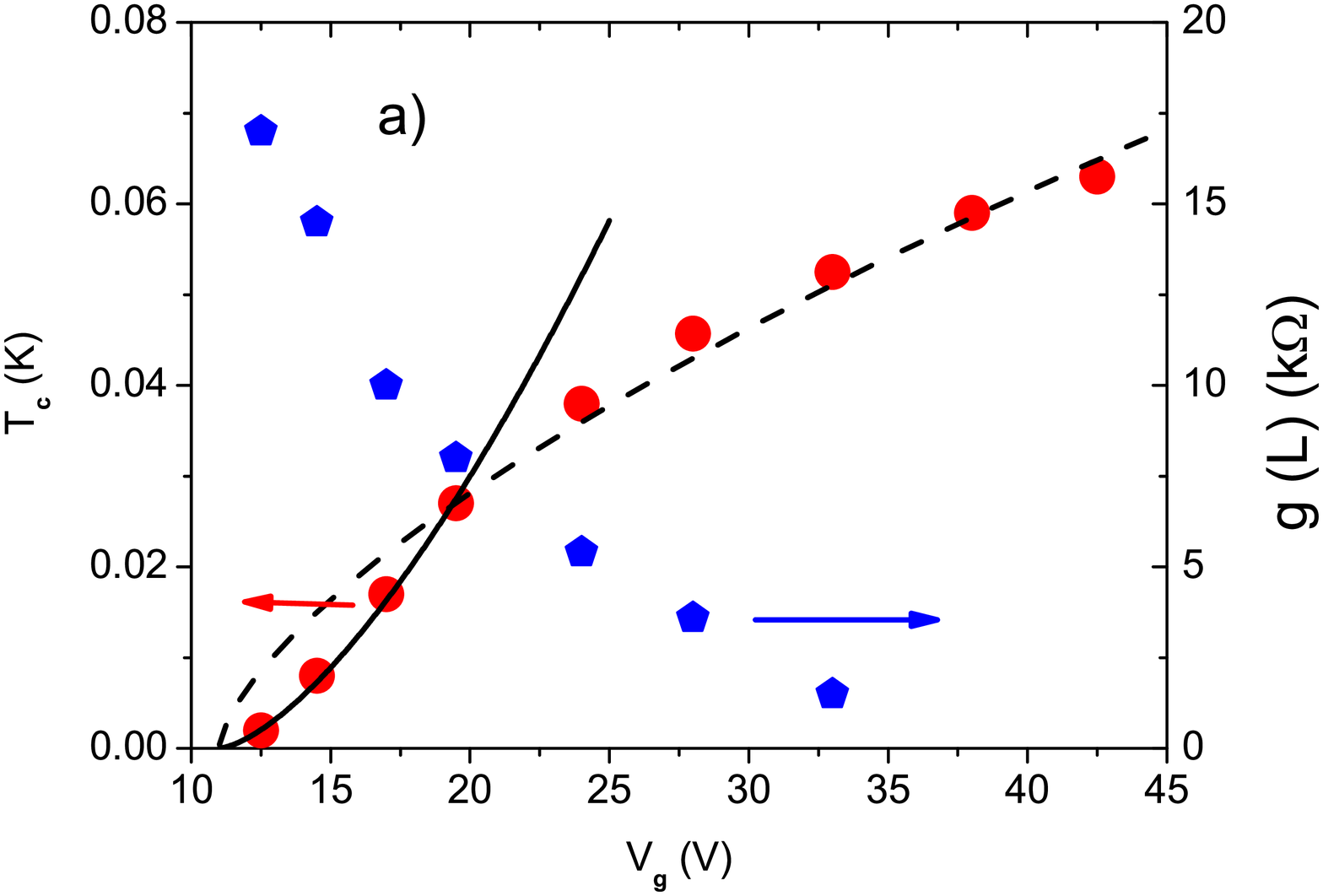}
\includegraphics[width=\linewidth]{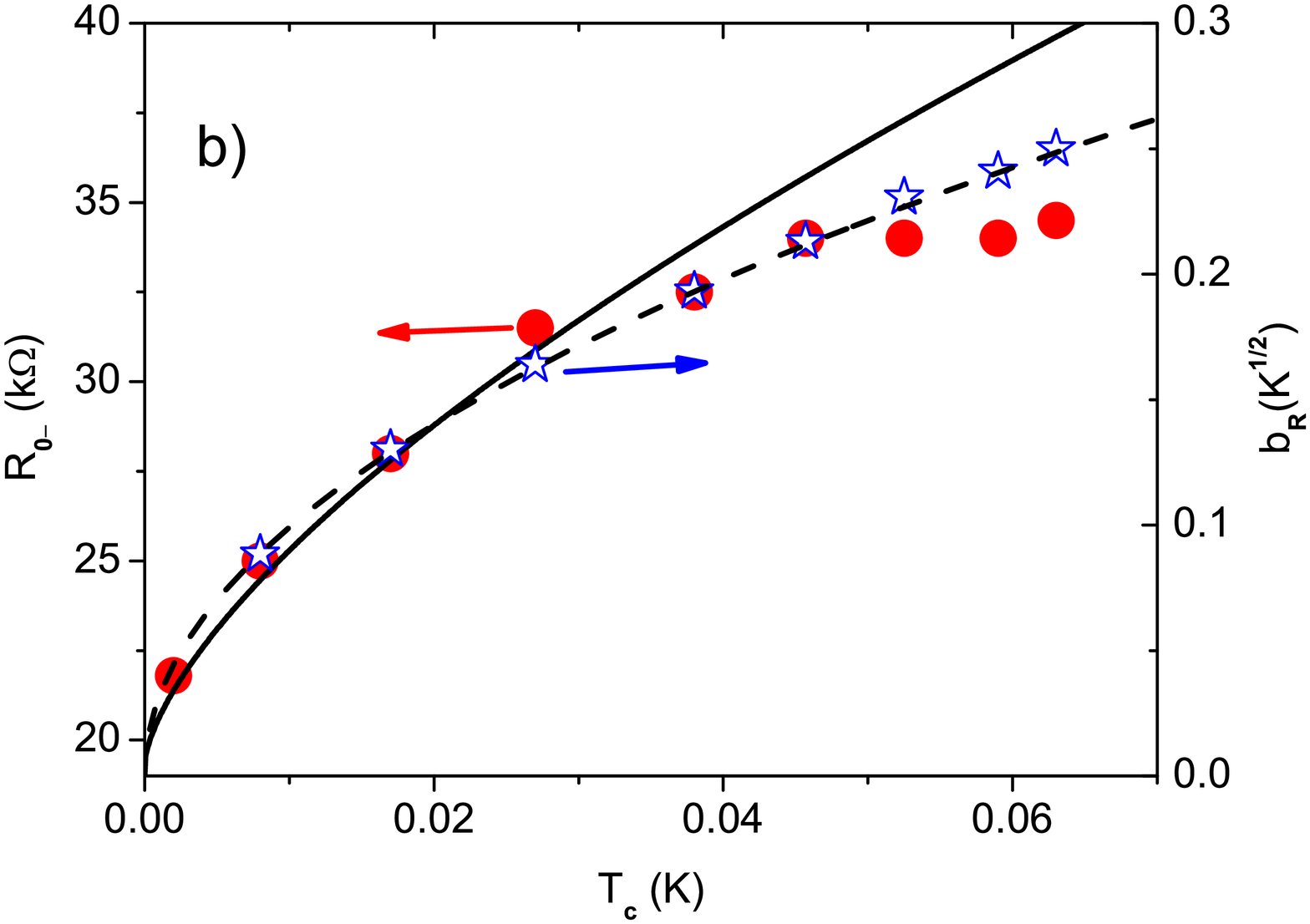}
\includegraphics[width=\linewidth]{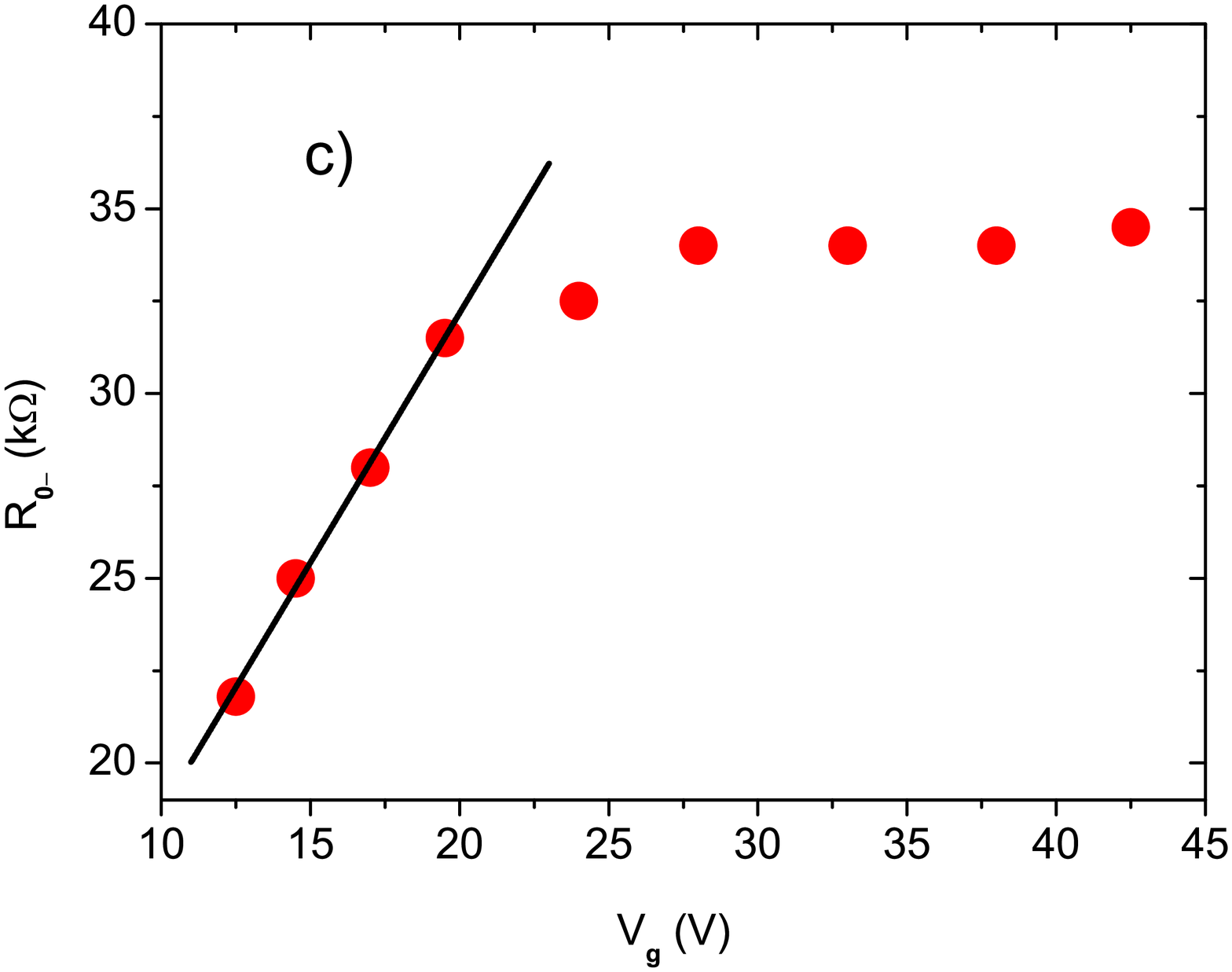}
\caption{Gate voltage and $T_{c}$ dependence of the BKT- and finite size scaling parameters $T_{c}$, $R_{0}$, $b_{R}$ and $g/L^{2}$. a) $T_{c}$ and $g$ \textit{vs}. $V_{g}$. The solid line is $T_{c}=0.0011~{\rm KV}^{-3/2}\left(V_{g}-V_{gc}\right)^{3/2}$ and the dashed one $T_{c}=0.0065~{\rm KV}^{-2/3}\left(V_{g}-V_{gc}\right) ^{2/3}$ with $V_{gc}=10.5$~V. b) $R_{0}$ and $b_{R}$ \textit{vs.} $T_{c}$. The solid line is $R_{0}=R_{0c}+128T_{c}^{2/3}$k${\rm \Omega} $ with $R_{0c}=19.35$ k${\rm \Omega} $ and the dashed one $b_{R}=0.99T_{c}^{1/2}$ K$^{1/2}$. c) $R_{0}$ \textit{vs}. $V_{g}$. The solid line is $R_{0}=R_{c}+1.4\left( V_{g}-V_{gc}\right) $k${\rm \Omega} $ with $V_{gc}=10.5$~V and $R_{c}=19.35$ k${\rm \Omega} $.}
\label{fig3}       
\end{figure}

The estimates for the BKT parameters and their gate voltage dependence are shown in Fig.~\ref{fig3}. Fig.~\ref{fig3}a uncovers a BKT-transition line $T_{c}\left( V_{g}\right) $ with a quantum critical endpoint around $V_{gc}=10.5$~V. Though these transition temperatures are not attained, the gate voltage dependent limiting lengths turned out to be sufficiently large to estimate them and the other parameters of interest from the intermediate temperature regime where consistency with BKT-behavior was established and confirmed in terms of a finite scaling analysis. The gate voltage dependence of $g\left( L\right) $ reveals that electrostatic tuning does not change the carrier density only but the inhomogeneity landscape as well. Indeed $g\left( L\right) \propto L^{-2}$ increases from $V_{g}=33$~V to $V_{g}=12.5$~V by almost an order of magnitude. Note that $g\left( L\right) $ fixes the width of the intermediate regime which shrinks with increasing $g\left( L\right) $, \textit{i.e.} with decreasing limiting length $L$.

The solid and dashed lines in Fig.~\ref{fig3}a are
\begin{equation}
T_{c}\left( V_{g}\right) =0.0011~{\rm KV}^{-3/2}\cdot\left( V_{g}-V_{gc}\right) ^{3/2},
\label{eq11}
\end{equation}
and
\begin{equation}
T_{c}\left( V_{g}\right) =0.0065~{\rm KV}^{-2/3}\left( V_{g}-V_{gc}\right) ^{2/3},
\label{eq12}
\end{equation}
with $V_{gc}=10.5$~V. They indicate limiting and effective power law behavior of the BKT phase transition line near its quantum critical endpoint where $T_{c}$ vanishes. In Fig.~\ref{fig3}b we show the estimates for the $T_{c}$ dependence of $R_{0-}$ and $b_{R}$. $R_{0-}$ is the amplitude of the BKT expression for the sheet resistance (Eq. (\ref{eq3})) and $b_{R}$ is related to the vortex core energy and should scale as \cite{tsprb,dahm,steel} 
\begin{equation}
b_{R}\left( V_{g}\right) =\widetilde{b}_{R}T_{c}^{1/2}\left( V_{g}\right) 
\label{eq13}
\end{equation}
A glance at Fig.~\ref{fig3}b reveals that this behavior is very well confirmed with the nonuniversal parameter
\begin{equation}
\widetilde{b}_{R}\simeq 0.99.  
\label{eq14}
\end{equation}
The estimates for $R_{0-}\left( V_{g}\right) $ shown in Figs.~~\ref{fig3}b and c also reveal that this amplitude exhibits a pronounced $T_{c}$ and gate voltage dependence. In the limits $T_{c}\rightarrow 0$ and $V_{g}\rightarrow V_{gc}\simeq 10.5$~V it tends to the critical value $R_{0c}\simeq 19.35$~k${\rm \Omega} $. The dashed line in Figs.~\ref{fig3}b is
\begin{equation}
R_{0-}\left( T_{c}\right) =R_{c}+128~{\rm k\Omega K}^{-2/3}\cdot T_{c}^{2/3} ,
\label{eq15}
\end{equation}
and the solid line in Fig.\ref{fig3}c 
\begin{equation}
R_{0-}\left( V_{g}\right) =R_{c}+1.35~{\rm k\Omega V}^{-1}\cdot \left( V_{g}-V_{gc}\right).  \label{eq16}
\end{equation}
They indicate power law behavior near the quantum critical point at $V_{gc}\simeq 10.5$~V and $R_{c}\simeq 19.35$~k${\rm \Omega} $ where $T_{c}$ vanishes. The reliability of these relationships can be checked by
eliminating $R_{0-}\left( V_{g}\right) -R_{c}$ and solving for $T_{c}(V_{g})$. This yields Eq. (\ref{eq11}) as it should be.

\begin{figure}
\includegraphics[width=\linewidth]{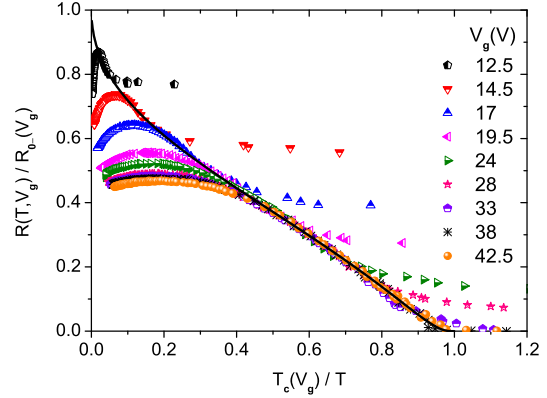}
\caption{Scaling plot $R(T,V_{g})/R_{0-}(V_{g})$ \textit{vs.} $T_{c}(V_{g})/T$ for the data shown in Fig. \ref{fig1} using the $T_{c}(V_{g})$ and $R_{0-}(V_{g})$ values shown in Fig.~\ref{fig3}a and Fig.~\ref{fig3}c, respectively. The solid line is $\exp\left( -\widetilde{b}_{R}/\left( T/T_{c}\left( V_{g}\right) -1\right)^{1/2}\right) $ in to Eq. (\ref{eq17}) with $\widetilde{b}_{R}\simeq 0.99$ (Eq. (\ref{eq14})).}
\label{fig4}       
\end{figure}

An essential implication of relation (\ref{eq13}) is that the BKT expression (\ref{eq3}) for the sheet resistance adopts the two variable scaling form 
\begin{eqnarray} 
\frac{R\left( T,V_{g}\right) }{R_{0-}\left( V_{g}\right) }&=&\left( \frac{\xi_{0}\left( V_{g}\right) }{\xi \left( T,V_{g}\right) }\right) ^{2} \nonumber\\
&=&\exp\left( -\frac{\widetilde{b}_{R}}{\left( T/T_{c}\left( V_{g}\right) -1\right)^{1/2}}\right). 
\label{eq17}
\end{eqnarray}
Indeed it depends on $R_{0-}\left( V_{g}\right) $ and $T_{c}\left(V_{g}\right) $ only. Accordingly, given $R_{0}\left( V_{g}\right) $ and $T_{c}\left( V_{g}\right) $ the sheet resistance data $R\left( T,V_{g}\right)$ plotted as $R\left( T,V_{g}\right) /R_{0-}\left( V_{g}\right) $ \textit{vs}. $T_{c}\left( V_{g}\right) /T$ should fall on a single curve given by the
right hand side of Eq. (\ref{eq17}). Deviations from this curve are expected close to $T_{c}\left( V_{g}\right) /T\lesssim 1$ due to the finite size effect and for $T_{c}\left( V_{g}\right) /T<<1$ where BKT fluctuations no longer dominate. This scaling plot is shown in Fig.~\ref{fig4}. The piecewise collapse of the data onto the scaling function, marked by the solid line clearly confirms that the BKT regime is attained. For large values of the scaling argument it is bounded by the finite size effect with a limiting length which increases with reduced gate voltage (Fig.~\ref{fig3}a) and for small arguments they reflect the fact BKT fluctuations no longer dominate.

\begin{figure}
\includegraphics[width=\linewidth]{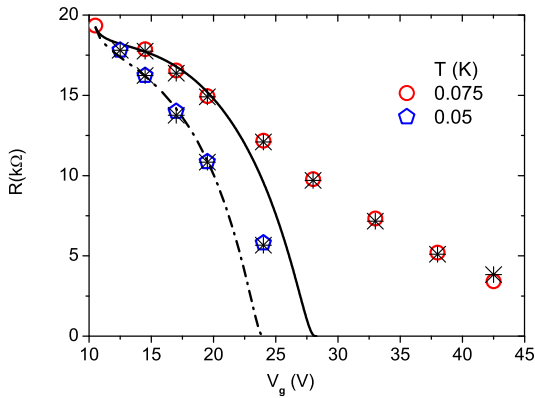}
\caption{Sheet resistance isotherms $R\left( V_{g}\right) $ at $T=0.075$~K and $T=0.05$~K. $\bigcirc $ and $\square $ denote the measured data points derived from Fig.~\ref{fig1}. $\ast $ mark the values derived from Eq. (\ref{eq17}) using the $T_{c}(V_{g})$ and $R_{0}(V_{g})$ values shown in Fig.~\ref{fig3}a and Fig.~\ref{fig3}c. The solid and dashed lines are derived from Eq. (\ref{eq17}) using the $T_{c}(V_{g})$ and $\ R_{0}(V_{g})$ given by the power laws (\ref{eq11}) and (\ref{eq16}).}
\label{fig5}       
\end{figure}

Evidence for a gate voltage driven QPT emerges from the phase transition line $T_{c}(V_{g})$ pointing to an endpoint where $T_{c}$ vanishes (Fig.~\ref{fig3}a) at a critical gate voltage $V_{gc}\simeq 10.5$~V and the amplitude $R_{0}\left( V_{g}\right) $ of the sheet resistance tends to the critical value $R_{0c}\simeq 19.5$~k${\rm \Omega} $ (Fig.~\ref{fig3}c). A characteristic property of a QPT at the endpoint of a BKT phase transition line follows from Eq. (\ref{eq17}) by considering the isotherms. Indeed, as $T_{c}\left( V_{g}\right)$ tends to zero by approaching $V_{gc}$ at fixed $T$, the amplitude of the sheet resistance $R(T,V_{g})$ tends to the temperature independent value $R_{c}$ at $V_{gc}$. Accordingly, the isotherms merge at the endpoint of a BKT transition line $T_{c}\left( V_{g}\right) $ which is a characteristic property of a QPT. In Fig.~\ref{fig5}, showing sheet resistance isotherms derived from the data depicted in Fig.~\ref{fig1}, we observe that the flow of the measured data points ( $\bigcirc $, $\square $) to a merging point near $V_{gc}\simeq10.5$~V and $R_{c}\simeq 19.35$~k${\rm \Omega} $ is well confirmed and consistent with respective values ($\star $) derived from Eq. (\ref{eq17}) and the $T_{c}(V_{g})$ and $R_{0}(V_{g})$ values shown in Fig.~\ref{fig3}a and Fig.~\ref{fig3}c. In contrast, the solid and dashed lines, obtained from Eq. (\ref{eq17}) and the power law expressions for $T_{c}(V_{g})$ and $\ R_{0}(V_{g})$ (Eqs. (\ref{eq11}) and (\ref{eq16})) apply closer to the QPT only.

\subsection{Quantum phase transition}
\label{sec:2:2}

The scaling theory of QPT yields for the sheet resistance to the one variable scaling form \cite{sondhi,fisher}
\begin{equation}
\frac{R\left( T,V_{g}\right) }{R_{c}}=G_{\pm }\left( y\right) ,y=\frac{c\left\vert V_{g}-V_{gc}\right\vert ^{z\overline{\nu }}}{T},  
\label{eq18}
\end{equation}
$c$ is a nonuniversal coefficient of proportionality, $R_{c}$ the critical sheet resistance, $z$ the dynamic and $\overline{\nu }$ the critical
exponent of the zero temperature correlation length, 
\begin{equation}
\xi \left( T=0,V_{g}\right) =\xi _{0}\left( T=0\right) \left\vert V_{g}-V_{gc}\right\vert ^{-\overline{\nu }}. 
\label{eq18a}
\end{equation}
$G_{\pm }\left( y\right) $ is a universal scaling function of its argument such that $G_{\pm }\left( y=0\right) =1$. The subscript $\left( +\right) $ marks the NI-crossover and $\left( -\right) $ the NS-transition line. Due to this line, corresponding in the present case to the BKT-transition line, the scaling function $G_{-}\left( y\right) $ exhibits at finite temperature a singularity at the universal vale $y_{c}$ of the scaling variable. The leading behavior of the phase transition line (\ref{eq3}) is then fixed by
\begin{equation}
T_{c}\left( V_{g}\right) =\frac{c}{y_{c}}\left\vert V_{g}-V_{gc}\right\vert^{z\overline{\nu }}.  
\label{eq19}
\end{equation}
Noting that in this limit
\begin{equation}
\frac{y}{y_{c}}=\frac{T_{c}\left( V_{g}\right) }{T},  
\label{eq20}
\end{equation}
the quantum scaling function $G_{-}\left( y\right) $ adopts according to Eq. (\ref{eq17}) the explicit BKT form
\begin{equation}
G_{-}\left( y\right) =\exp \left( -\widetilde{b}_{R}\left( y_{c}/y-1\right)^{-1/2}\right).
\label{eq21}
\end{equation}
Because $R_{c}$ is fixed there is one scaling parameter only, namely $y\propto T_{c}\left( V_{g}\right) /T$. \ Taking the gate voltage dependence of $R_{0}$ into account we obtain the two parameter scaling form
\begin{equation}
\frac{R\left( T,V_{g}\right) }{R_{0-}\left( V_{g}\right) }=G_{-}\left(y\right).
\label{eq22}
\end{equation}

\begin{figure}
\includegraphics[width=\linewidth]{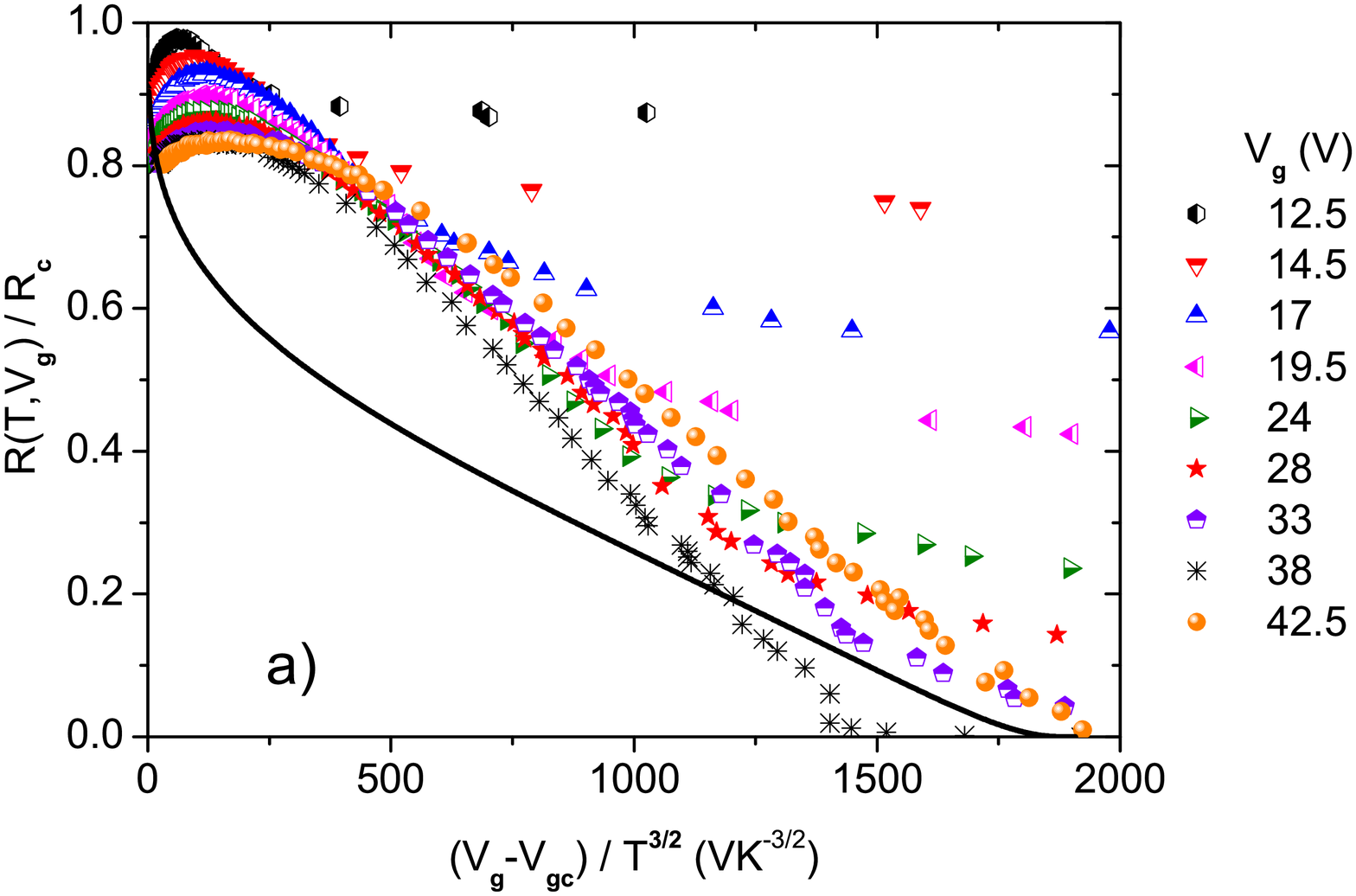}
\includegraphics[width=\linewidth]{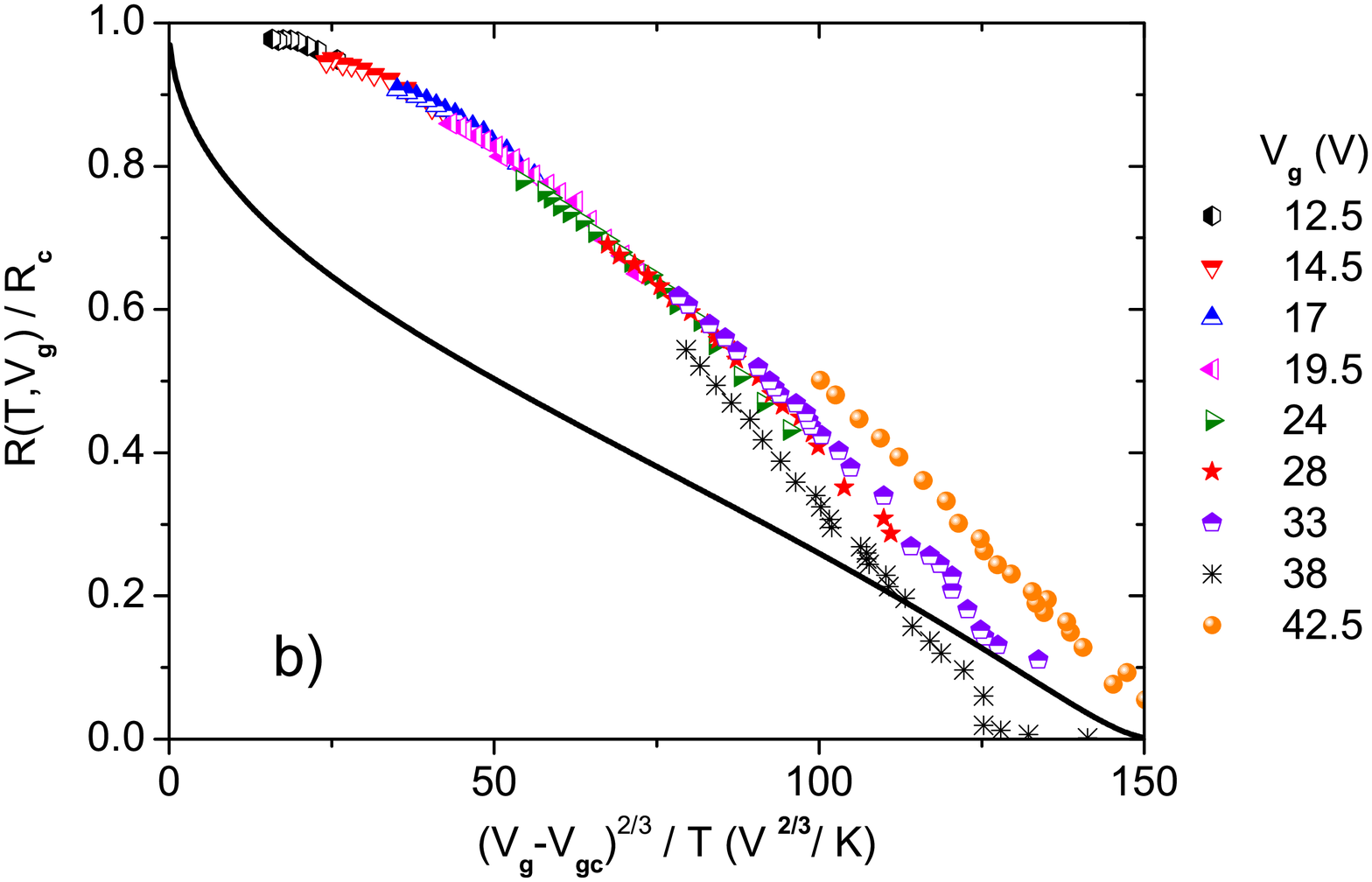}
\caption{Quantum scaling plot $R\left( T,V_{g}\right) /R_{c}$ $vs.$ $\left(V_{g}-V_{g}c\right) ^{z\overline{\nu }}/T$ with $z\overline{\nu }=2/3$, $T_{c}=19.5$~k${\rm \Omega} $ and $V_{gc}=10.5$~V. a) For the data shown in Fig.~\ref{fig1}. The solid line is Eq. (\ref{eq17}) with $R_{0}(V_{g})=R_{0}(V_{gc})=R_{c}=19.35$~k${\rm \Omega} ,\widetilde{b}_{R}=0.99$, and $T_{c}\left( V_{g}\right) =0.0065~{\rm KV}^{-2/3} \left( V_{g}-V_{gc}\right) ^{z\overline{\nu }}$ with $z\overline{\nu }=2/3$ (Eqs. (\ref{eq12}) and (\ref{eq19})) indicates the BKT behavior. b) For the same data but restricted to $0.06$~K$<T<0.1$~K and the same solid line.}
\label{fig6}       
\end{figure}

The occurrence of a QPT at the endpoint of the BKT-transition line implies that the isotherms merge at the critical value of the tuning parameter $V_{gc}$ with sheet resistance $R_{c}$. where at $T=0$ the QSI-transition occurs. In the BKT scenario where the NS-branch is considered this crossing point corresponds to the merging point seen in Fig.~\ref{fig5}. In the quantum scaling approach the scaled data $R\left( T,V_{g}\right) /R_{c}$ \textit{vs}. $\left\vert V_{g}-V_{gc}\right\vert ^{z\overline{\nu }}/T$ should then fall onto two branches crossing at the QSI critical point. The upper branch stems from the NI-crossover and the lower one from the NS- transition. $z\overline{\nu }$ is usually determined by minimizing the error in data collapse \cite{markovic,gant,parendo,parendo2}. The dashed line in Fig.~\ref{fig3}a, given by Eq. (\ref{eq12}), suggests that the effective exponent $z\overline{\nu }\simeq 2/3$ mimics the overall gate voltage dependence of $T_{c}$ reasonably well. The resulting scaling plot, including the data shown in Fig.~\ref{fig1} is depicted in Fig.~~\ref{fig6}a. Admittedly the data collapse is rather poor. For large values of the scaling argument the collapse is affected by the finite size effect with a limiting length which increases with reduced gate voltage (Fig.~\ref{fig3}a) and for small values of the scaling argument one enters $T>>$ $T_{c}(V_{g})$ where BKT fluctuations no longer dominate. More alarming are the large deviations from the BKT-behavior indicated by the solid line. This discrepancy is attributable to the fact that at this gate voltages the amplitude $R_{0-}$ decreases substantially and differs considerably from the critical value $R_{c}$. On the other hand adopting the suggestion of Parendo \textit{et al.} \cite{parendo,parendo2} the quality of the data collapse can be improved by restricting the data, \textit{i.e.} to the interval $0.06$~K $<T<0.1$~K. As shown in Fig.~\ref{fig6}b this improves the quality of the collapse considerably and suggests consistency with $z\overline{\nu }\simeq 2/3$. Although on a first glance compelling, this procedure is misleading because the asymptotic regime where the quantum scaling form (\ref{eq18}) applies is not attained as the gate voltage dependence of $T_{c}$ (see Fig.~\ref{fig3}a) and the amplitude $R_{0-}\left( V_{g}\right) $ reveals (Fig.~\ref{fig3}c).

\begin{figure}
\includegraphics[width=\linewidth]{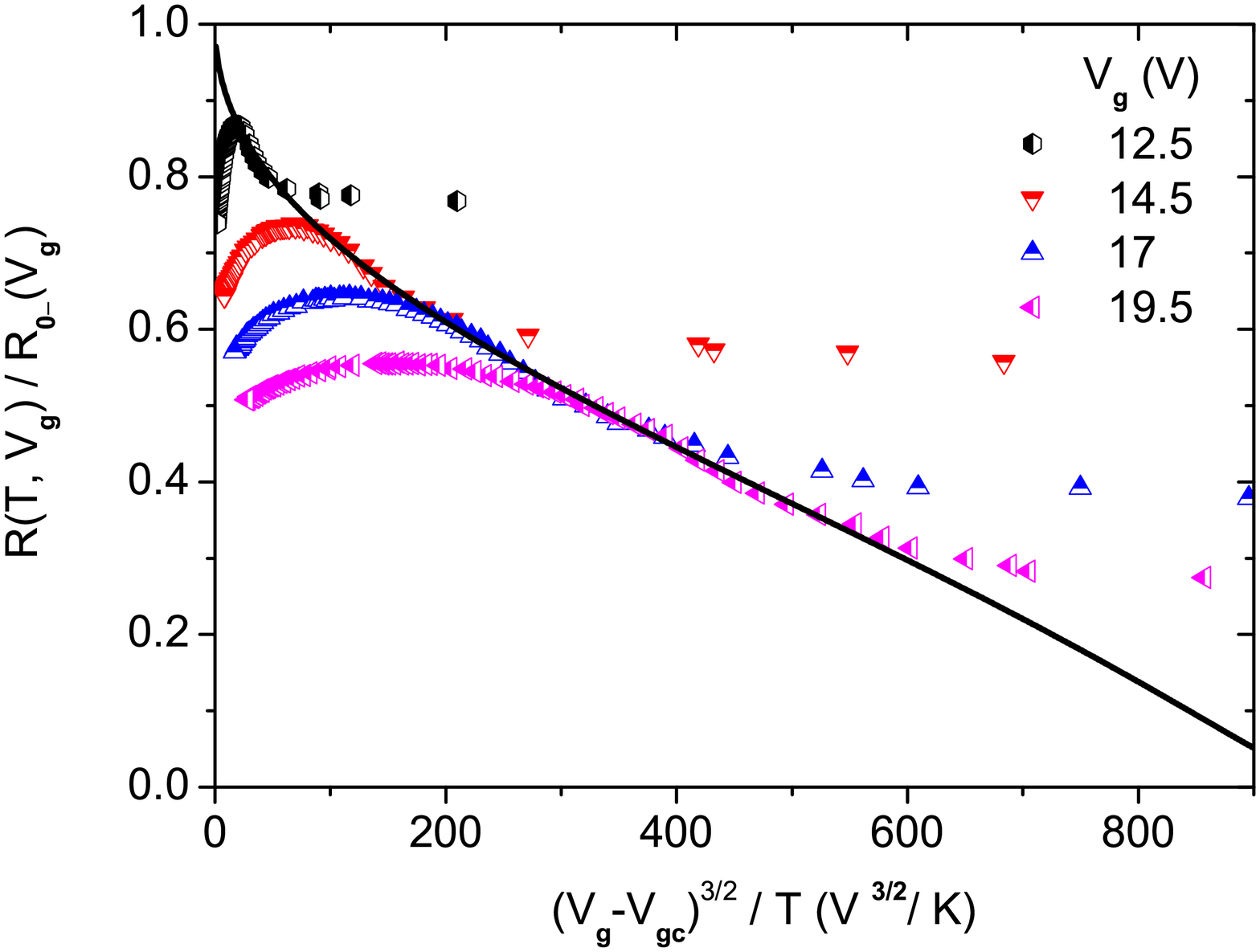}
\caption{Scaling plot $R\left( T,V_{g}\right) /R_{0-}\left( V_{g}\right) $ $vs.$ $\left( V_{g}-V_{gc}\right) ^{z\overline{\nu }}/T$ with $z\overline{\nu }=3/2$, $T_{c}=19.5$~k${\rm \Omega} $, $V_{gc}=10.5$~V and The solid line is the BKT scaling function Eq. (\ref{eq21}) \ with $\widetilde{b}_{R}=0.99$, $T_{c}\left( V_{g}\right) =0.0011~{\rm KV}^{-3/2}\left( V_{g}-V_{gc}\right) ^{z\overline{\nu }}$ and $z\overline{\nu }=3/2$ (Eqs. (\ref{eq11}) and (\ref{eq19})). $R_{0}= $ $R_{0-}(V_{g})$ is taken from Fig.~\ref{fig3}c.}
\label{fig7}       
\end{figure}

Indeed, Fig. ~\ref{fig3}a shows that $z\overline{\nu }\simeq 2/3$ is at best an effective exponent failing near criticality where $z\overline{\nu }$ certainly exceeds one and $z\overline{\nu }\simeq 3/2$ is consistent with the available data. Furthermore $z\overline{\nu }\simeq 3/2$ is consistent with the limiting behavior of the sheet resistance isotherms close to the quantum critical point as depicted in Fig.~\ref{fig5}. According to Fig.~\ref{fig7} this value also leads to a piecewise excellent data collapse onto the BKT scaling function (\ref{eq21}) using the two parameter scaling form (\ref{eq22}). For large values of the scaling argument the deviations stem from the finite size effect with a limiting length which increases with reduced gate voltage (Fig.~\ref{fig3}a) and for small arguments they reflect the fact that BKT fluctuations no longer dominate. Considering Fig.~\ref{fig3}a we observe that the attainable quantum critical regime extends up to $V_{g}\simeq 19.5$~V and according to Fig.~\ref{fig2}b up to the temperature $T\simeq 0.07$~K. A restriction of the sheet resistance data, \textit{i.e.} to the interval $0.06$~K $<T<0.1$~K implies then that in the scaling plot shown in Fig.~\ref{fig6}b an essential part of the quantum critical regime is missing.

Next we turn to the gate voltage and $T_{c}$ dependence oft the BKT amplitude $R_{0}$ shown in Figs.~\ref{fig4}b and \ref{fig4}c. In $D=2$ the sheet resistance, while diverging in the insulating phase, and zero in the superconducting phase, has a finite constant value $R_{c}$ right at the quantum critical point. Indeed along the BKT phase transition line approach both $R_{0-}(V_{g})$ and $R_{0-}(T_{c})$ the critical value $R_{c}\simeq 19.5$~k${\rm \Omega} $. We observe that In the quantum critical regime outlined above ($V_{g}\lesssim 19.5$~V and $T_{c}\lesssim 0.07$~K) the approach to $R_{c}$ is consistent with the power laws (\ref{eq15}) and (\ref{eq16}). Note that in
analogy to $b_{R}\left( V_{g}\right) $ and $T_{c}\left( V_{g}\right) $ the amplitude $R_{0-}(V_{g})$ is also renormalized by quantum fluctuations. To estimate the exponents $z$ and $\overline{\nu }$ we note that according to classical dynamic scaling $R_{0-}\left( V_{g}\right) $ is proportional to $\xi _{0}^{-2}\left( V_{g}\right) $ where $\xi _{0}\left( V_{g}\right) $ is the amplitude of the BKT correlation length 
\begin{eqnarray}
& {\rm } & \xi \left( T,V_{g}\right) = \nonumber \\
& {\rm } & \xi _{0}\left( V_{g}\right) {\rm exp}\left(-\left( \widetilde{b}_{R}/2\right) \left( T/T_{c}\left( V_{g}\right)-1\right) ^{-1/2}\right) .  
\label{eq22a}
\end{eqnarray}
At low temperatures where $T_{c}\left( V_{g}\right) \propto \left(V_{g}-V_{gc}\right) ^{z\overline{\nu }}$ (Eq. (\ref{eq19}) ) applies it has to match the quantum scaling counterpart 
\begin{eqnarray}
\xi \left( T,V_{g}\right) &\propto& \left( V_{g}-V_{gc}\right) ^{-\overline{\nu }}f_{-}\left( y\right) \nonumber \\
y&=&cT/\left( V_{g}-V_{gc}\right) ^{z\overline{\nu}}\propto T/T_{c}, 
\label{eq22b}
\end{eqnarray}
extending Eq. (\ref{eq18a}) to finite temperatures. The scaling function $f_{-}\left( y\right)$ adopts the limits:
\begin{equation}
f_{-}\left( y\right) =\left\{
\begin{array}{c}
1~:y=0 \\ 
y^{-\overline{\nu }}:y=\infty
\end{array}
\right. 
\label{eq22c}
\end{equation}
The matching yields an explicit form of the scaling function $f_{-}\left(x\right) $ and reveals that $\xi _{0}\left( V_{g}\right) $ scales as $\xi_{0}\left( V_{g}\right) =\xi \left( T=0,V_{g}\right) \propto \left(V_{g}-V_{gc}\right) ^{-\overline{\nu }}\propto T^{-1/z}$. \ On the other hand there is the quantum constraint of a finite critical sheet resistance $R_{c}$ at $T=0$ which is incompatible with the classical dynamic scaling prediction $R_{0-}\left( V_{g}\right) $ $\propto $ $\xi _{0}^{-2}\left(V_{g}\right) $ because $\xi _{0}^{-2}\left( V_{g}\right) $ tends to zero as $\left( V_{g}-V_{gc}\right) ^{2\overline{\nu }}$. The quantum constraint is then accounted for in terms of \cite{tsprb,tssw}
\begin{eqnarray}
R_{0-}(V_{g})-R_{c}&\propto& 1/\xi ^{2}\left( T=0,V_{g}\right) \nonumber\\
&\propto& \left(V_{g}-V_{gc}\right) ^{2\overline{\nu }}\propto T_{c}^{2/z}.  
\label{eq24}
\end{eqnarray}
Using the estimates for $R_{0-}(V_{g})$, $R_{c}$ and $T_{c}\left(V_{g}\right) $ resulting from the finite size scaling analysis of the temperature dependence of the sheet resistance at fixed gate voltages this relationship provides a direct way to measure the critical exponents $z$ and $\overline{\nu }$, and allows for a check on the value of $z\overline{\nu }$. Previously it was confirmed in the analysis of the sheet resistance data of the superconducting LaAlO$_{3}$/SrTiO$_{3}$ interface \cite{tsprb} and thin La$_{2-x}$Sr$_{x}$CuO$_{4}$ films \cite{tssw}. Furthermore $\xi_{0}\propto T_{c}^{-1/z}$ was verified in submonolayer superfluid $^{4}$He films \cite{williams}. Combining Eqs. (\ref{eq24}), (\ref{eq15}) and (\ref{eq16}) we obtain for the critical exponents of the QSI transition the estimates 
\begin{equation}
z\overline{\nu }\simeq 3/2,z\simeq 3,\overline{\nu }\simeq 1/2.  
\label{eq25}
\end{equation}
Noting that $D+z=2+z\simeq 5$ exceeds the upper critical dimension $D_{u}=4$ the critical exponent of the zero temperature correlation length should adopt its mean-field value $\overline{\nu }=1/2$. However the fate of this clean QSI critical point under the influence of disorder is controlled by the Harris criterion \cite{harris,aharony}. If the inequality $\overline{\nu }\geq 2/D$ is fulfilled, the disorder does not affect the critical behavior. If the Harris criterion is violated ($\overline{\nu }<2/D$), the generic result is a new critical point with conventional power law scaling but new exponents which fulfill $\overline{\nu }<2/D$ and $\overline{\nu }\neq 1/2$. Because disorder is relevant it drives the system from the mean-field to an other critical point with different critical exponents and with that to $\overline{\nu }\neq 1/2$. Unfortunately the available data are too sparse to derive a more precise estimate of $\overline{\nu }$. Otherwise it is clear that $D+z=2+z\simeq 5$ exceeds the upper critical dimension $D_{u}=4$ so
that the the equivalence between quantum phase transitions in systems with $D $ spatial dimensions and the ones of classical phase transitions in $(D+z)$ dimensions does not apply.

\subsection{Quantum superconductor to insulator transition}
\label{sec:2:3}

In disordered 2D systems weak localization manifests itself due to the wave nature of electrons since interference effects double the classical probability for self-crossing paths and it slightly suppresses the conductance \cite{lee}. In the presence of strong spin-orbit coupling, the spins rotate in the opposite direction and the interference is destructive, which leads to an increase in the conductance, this effect being opposite to the weak localization and known as weak anti-localization \cite{bergmann}. The influence of spin-orbit coupling increases for heavy elements since it strongly depends on the atomic number Z, thus bismuth, being the heaviest group-V semimetal, is a suitable material for the study of its electronic transport properties through the analysis of the weak anti-localization contribution \cite{marcano}. The fermionic part comprises the sheet conductance and its modifications due to weak localization, electron-electron- and spin orbit interaction, yielding\cite{lachlan}
\begin{equation}
\sigma _{n}=\sigma _{1}+\sigma _{2}\ln \left( T\right).   
\label{eq26}
\end{equation}
Note that
\begin{equation}
\sigma _{2}=\frac{e^{2}}{\pi h}\simeq 1.23\times 10^{-2}~{\rm k}{\rm \Omega} ^{-1}
\label{eq27}
\end{equation} 
is generally attributed to the electron-electron interaction contribution \cite{blanter}. In the vicinity of the normal state to superconductor transition there is also a bosonic contribution due to Cooper pair fluctuations. In the Gaussian approximation it is given by the Aslamosov-Larkin (AL) expression (\ref{eq9}). In Fig.~\ref{fig8} we show the sheet
conductance $\sigma $ \textit{vs}. ln$(T)$ for gate voltages covering the normal state to superconductor transition and the normal state to insulator crossover. Considering the behavior at $V_{g}=0$~V and $V_{g}=7$~V we observe consistency with weak anti-localization. Indeed the gate voltage dependent values for $\sigma _{2}$ are substantially smaller than the electron-electron interaction contribution \cite{blanter}. The saturation above ln$(T)\approx -3$ is again attributable to the finite size effect preventing the diverging length associated with weak localization to grow beyond the limiting length $L$. Although there is evidence for weak anti-localization at $V_{g}=12.5$~V and even at $V_{g}=11$~V below ln$\left(T\right) \approx -2$ Cooper pair fluctuations dominate at low temperatures as the comparison with the AL- behavior (Eq. (\ref{eq9})) and BKT behavior reveals (see Fig.\ref{fig2}b). Here the approach to a normal state to superconductor transition, limited by the finite size outlined above, is apparent. Noting that at $T_{c}$, $\sigma \propto \xi ^{2}\propto L^{2}$, the gate voltage dependence of the saturation in $\sigma $ reflects the reduction of the limiting length $L$ with reduced gate voltage (see Fig.\ref{fig3}a). Similarly, as the gate voltage increases on the insulating side from $0$~V to $7$~V the saturation level decreases because the diverging length is limited again by a length which decreases by approaching the critical gate voltage $V_{g}\simeq 10.5$~V from below. The features emerging from Fig.~\ref{fig1} can be summarized as follows: A characteristic feature of the normal state appears to be the competition between weak anti-localization and superconductivity. For $V_{g}<10.5$~V superconductivity is suppressed and a finite size limited insulating ground state is approached. This approach is in agreement with previous work \cite{marcano} consistent with weak anti-localization. On the contrary for $V_{g}>10.5$~V pair fluctuations dominate at sufficiently low temperatures and a finite size limited approach to the superconducting ground state is observed. From the merging point seen in Fig.~\ref{fig5} we know that the flows to the superconducting and insulating ground states are separated by the the quantum critical point at $V_{gc}\simeq 10.5$~V and $\sigma_{c}=1/R_{c}\simeq 0.0517$~k${\rm \Omega} ^{-1}$. The temperature dependence of the sheet conductance at various gate voltages seen in Fig.~{\ref{fig8}a is then consistent with the approach to the QSI-critical point with finite critical sheet conductivity.

\begin{figure}
\includegraphics[width=\linewidth]{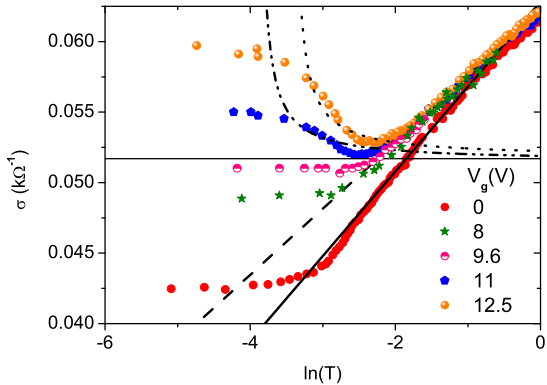}
\caption{Sheet conductance $\sigma \left( T,V_{g}\right) $ \textit{vs.} ln$(T)$ for data shown in Fig.~\ref{fig1}. The horizontal thin line marks the critical sheet conductance $\sigma _{c}=1/R_{c}\simeq 0.0517$~k${\rm \Omega} ^{-1}$. The dashed line is Eq. (\ref{eq26}) describing the fermionic contribution with $\sigma _{1}=0.062$~k${\rm \Omega} ^{-1}$ and $\sigma _{2}=4.69\times 10^{-3}$~k${\rm \Omega} ^{-1}$and the solid one with $\sigma _{1}=0.063$~k${\rm \Omega} ^{-1}$ and $\sigma _{2}=6.01\times 10^{-3}$~k${\rm \Omega} ^{-1}$. The dotted and dash-dot-dot lines are the AL expression (\ref{eq9}) describing the gaussian bosonic contribution to the normal state to superconductor transition with $\sigma_{n}=0.0514$~(k${\rm \Omega} ^{-1}$), $T_{c0}=0.036$~K, $\sigma _{n}=0.0505$~k${\rm \Omega} ^{-1}$ and $T_{c0}=0.028$~K. The dash-dot line is the BKT behavior resulting from Eq. (\ref{eq17}) with $\widetilde{b}_{R}=0.99$ (Eq. (\ref{eq14})), $R_{0-}=22.05$~k${\rm \Omega} $ and $T_{c}=0.002$~K given in Fig.~\ref{fig3}.}
\label{fig8}       
\end{figure}

Otherwise it is also expected that in disordered films variable-range hopping controls the insulating phase whereupon the sheet conductivity scales as \cite{mott} 
\begin{equation}
\sigma \left( T,V_{g}\right) =\sigma _{h}\left( V_{g}\right) \exp \left(-\left( T_{0}\left( V_{g}\right) /T\right) ^{1/3}\right) .  
\label{eq28}
\end{equation}
$T_{0}\left( V_{g}\right) $ denotes the variable-range hopping crossover temperature. In this context it should be kept in mind that the irrelevance of disorder at the BKT-transition does not exclude its relevance at zero temperature. As our estimate $\overline{\nu }\simeq 1/2$ violates in $D=2$ the Harris criterion we know that disorder is relevant at zero temperature. For this reason we consider below an insulating ground state controlled by variable-range hopping. An approach to the QSI critical point requires then that $\sigma _{h}\left( V_{g}\right) \rightarrow \sigma _{h}\left( V_{gc}\right) =\sigma _{c}$ and $T_{0}\left( V_{g}\right) \rightarrow T_{0}\left( V_{gc}\right) =0$. To explore this scenario quantitatively we note that Eq. (\ref{eq28}) leads according to Eq. (\ref{eq18}) to the quantum scaling form
\begin{equation}
\frac{R\left( T,V_{g}\right) }{R_{0+}\left( V_{g}\right) }=\frac{\sigma_{0+}\left( V_{g}\right) }{\sigma \left( V_{g}\right) }=G_{+}\left( y\right),y=\frac{c\left\vert V_{g}-V_{gc}\right\vert ^{z\overline{\nu }}}{T},
\label{eq29}
\end{equation}
with the explicit scaling function
\begin{equation}
G_{+}\left( y\right) \simeq \exp (fy^{1/3}),  
\label{eq30}
\end{equation}
and the crossover temperature
\begin{equation}
T_{0}\left( V_{g}\right) =f^{3}\left\vert V_{g}-V_{gc}\right\vert ^{z\overline{\nu }},  
\label{eq31}
\end{equation}
where the value of $z\overline{\nu }$ applies on both side of the QSI-transition. In Fig.~\ref{fig9}a we depicted the scaling plot \linebreak $R\left(V_{g}\right) /R_{0+}\left( V_{g}\right)$ \textit{vs.} $\left( T_{0}\left(V_{g}\right) /T\right) ^{1/3}$ by choosing \linebreak $R_{0+}\left( V_{g}\right)=1/\sigma _{h}\left( V_{g}\right) $ and $T_{0}\left( V_{g}\right) $ to minimize the error in data collapse onto the scaling function \linebreak exp$\left(T_{0}\left( V_{g}\right) /T\right) ^{1/3}$. Though the quality of the collapse is satisfactory and points to a variable-range hopping controlled insulating ground state, it is again subjected to the limiting length identified with the finite size scaling analysis of the BKT transition. It
leads to the saturation at low temperature. Additional insight emerges from the gate voltage dependence of the adjusted parameters $T_{0}\left(V_{g}\right) $ and $R_{0+}\left( V_{g}\right) $ shown in Figs.~\ref{fig9}b and \ref{fig9}c. In contrast to the BKT-transition line (see Fig.~\ref{fig3}a) the quantum scaling form (\ref{eq31}) of the crossover temperature $T_{0}\left( V_{g}\right) $ applies over the entire gate voltage range and confirms $z\overline{\nu }\simeq 3/2$. The strong gate voltage dependence of the amplitude $R_{0+}\left(V_{g}\right) $ uncovers again the limitatios of one parameter scaling. It appears to be related to the inverse of the zero temperature correlation length. Indeed close to the QSI transition we observe consistency with
\begin{equation}
R_{0+}\left( V_{g}\right) -R_{c}\propto \left\vert V_{g}-V_{gc}\right\vert^{1/2}\propto T_{0}^{1/3}.  
\label{eq32}
\end{equation}
It suggests that 
\begin{equation}
R_{0+}\left( V_{g}\right) -R_{c}\propto 1/\xi \left( T=0\right) \propto \left\vert V_{g}-V_{gc}\right\vert ^{\overline{\nu }}\propto T^{1/z},
\label{eq33}
\end{equation}
with $\overline{\nu }\simeq 1/2$ and $z\simeq 3$ applies, and with it these critical exponents apply on both sides of the QSI transition (see Eq. (\ref{eq25})).

\begin{figure}
\includegraphics[width=\linewidth]{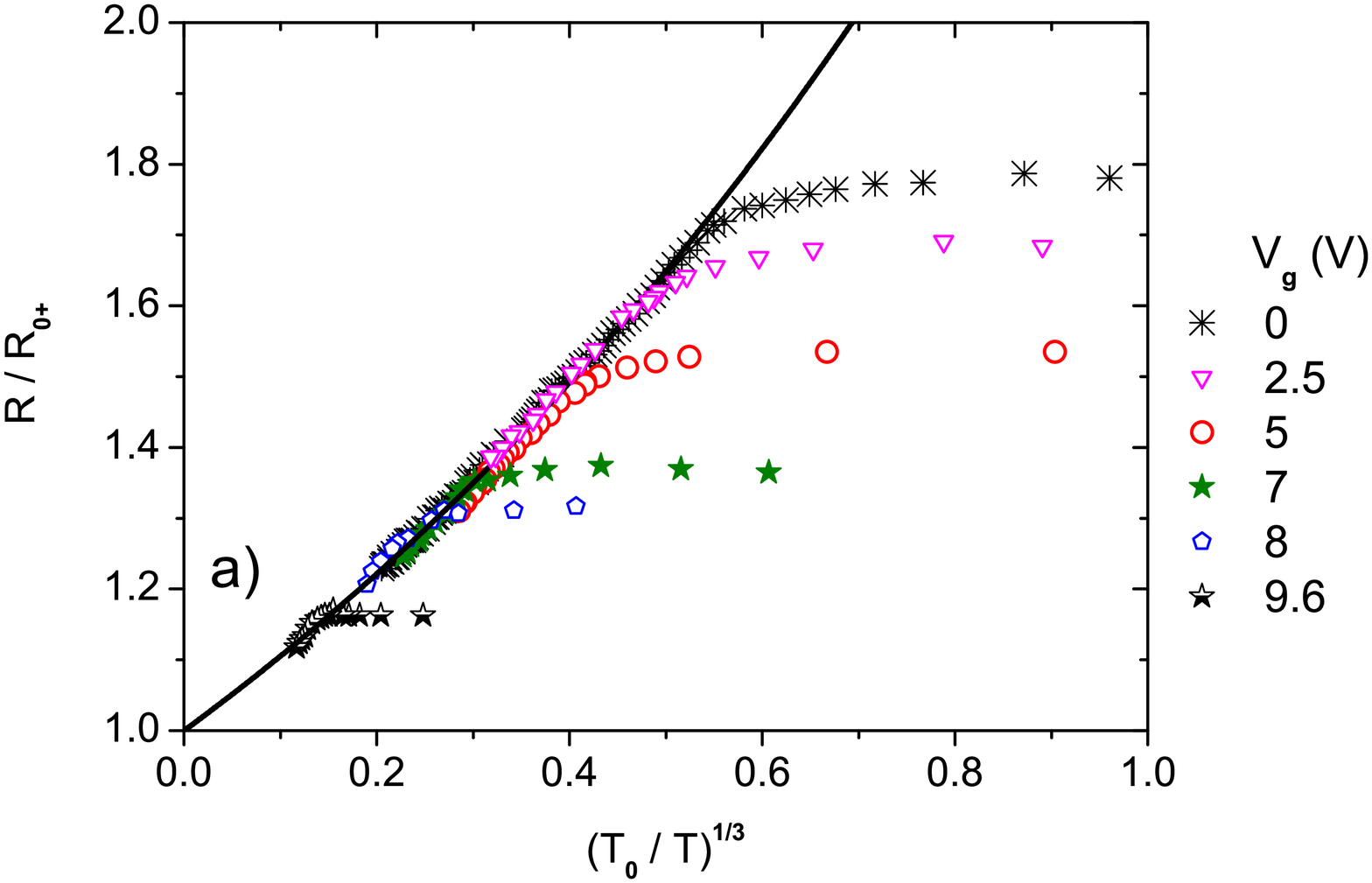}
\includegraphics[width=\linewidth]{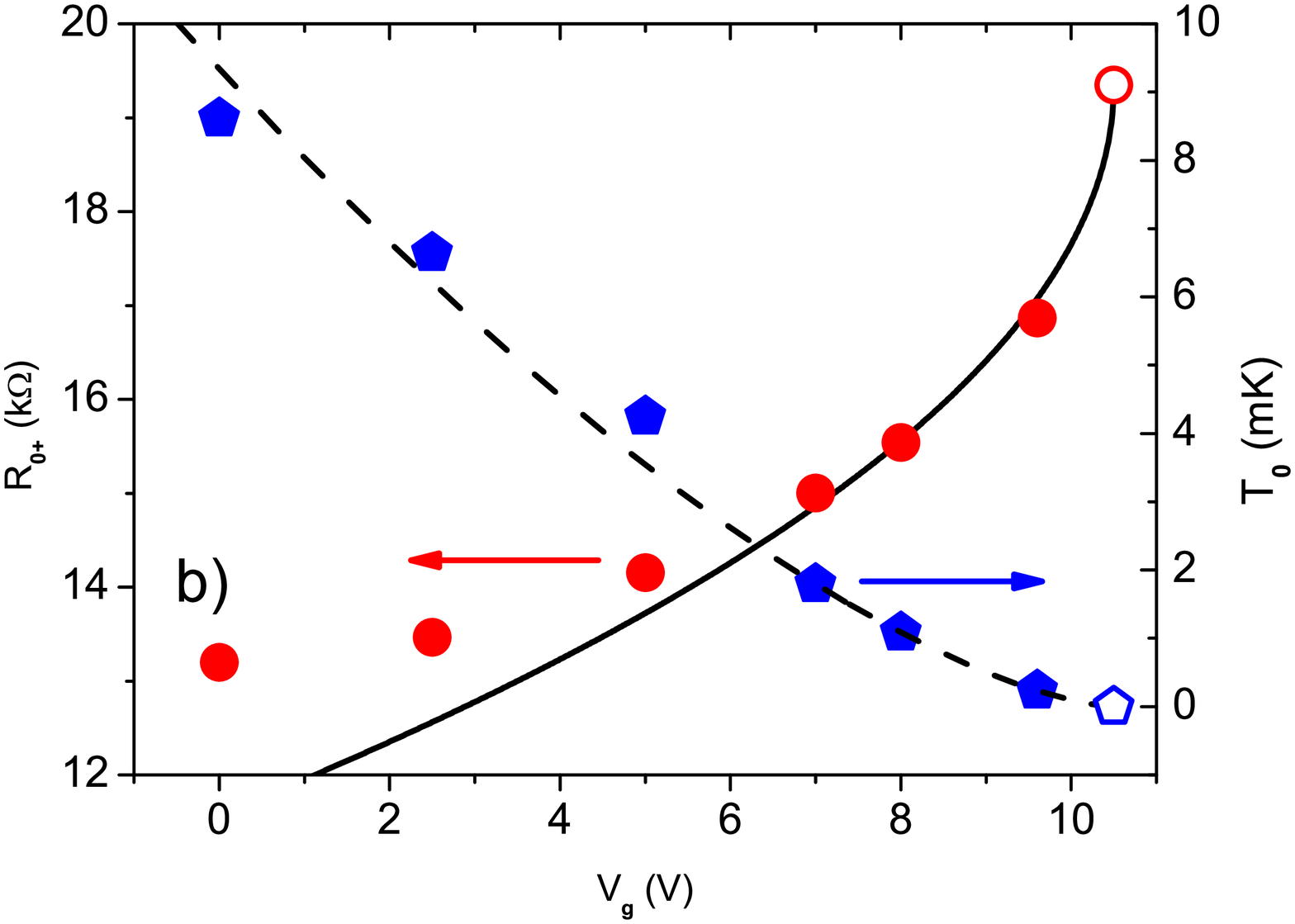}
\includegraphics[width=\linewidth]{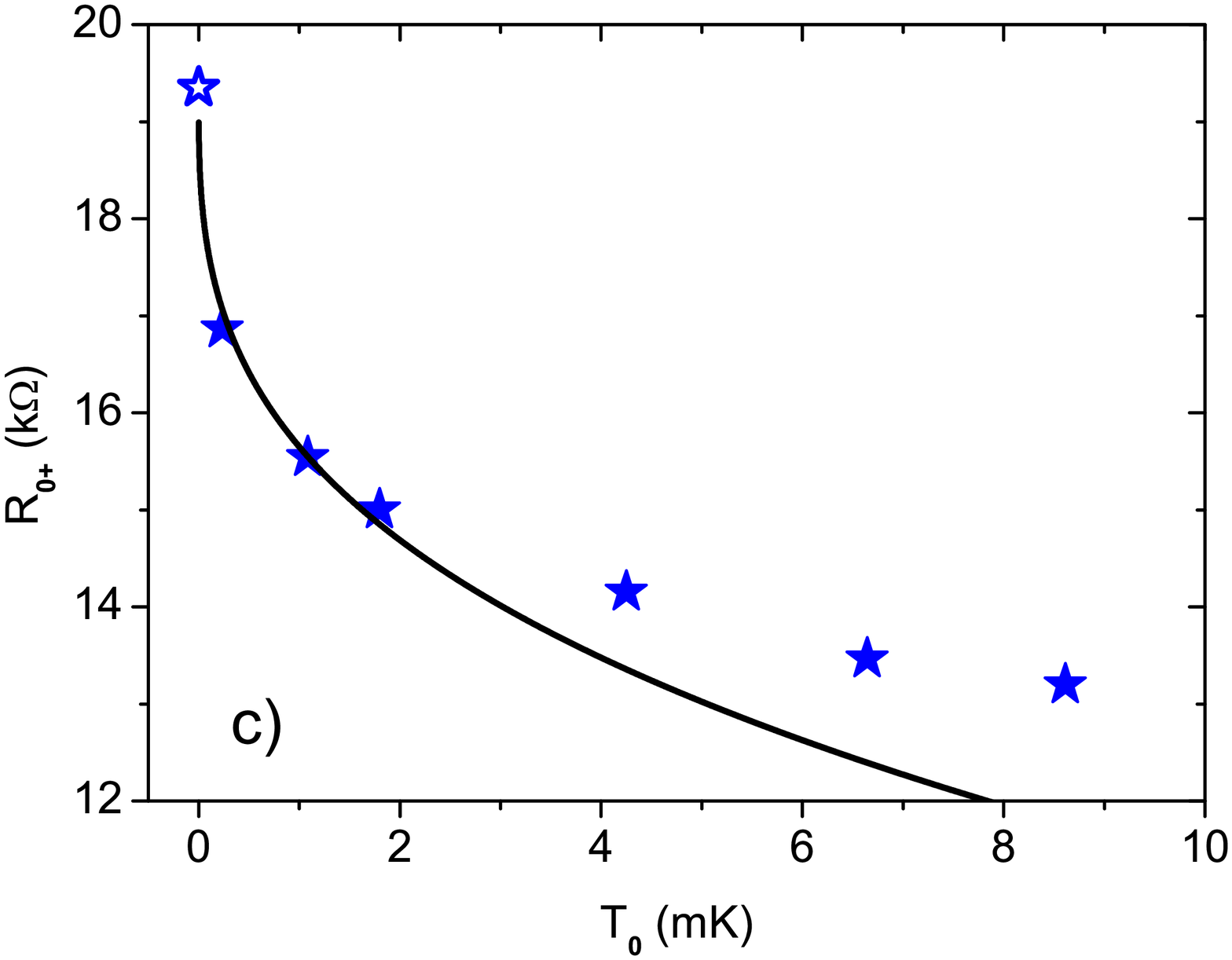}
\caption{a) $R\left( V_{g}\right) /R_{0+}\left( V_{g}\right) $ \textit{vs.} $\left( T_{0}\left( V_{g}\right) /T\right) ^{1/3}$ derived from the sheet resistance data of Parendo \textit{et al.} \cite{parendo,parendo2}. $R_{0+}\left( V_{g}\right) $ and $T_{0}\left( V_{g}\right) $ are chosen to minimize the error in data collapse onto the scaling function exp$\left(T_{0}\left( V_{g}\right) /T\right) ^{1/3}$. b) Resulting estimates for $R_{0+}\left( V_{g}\right) $ and $T_{0}\left( V_{g}\right) $. The dashed line is $T_{0}\left( V_{g}\right) \simeq f^{3}\left\vert V_{g}-V_{gc}\right\vert^{z\overline{\nu }}$ with $f=0.065~{\rm K}^{-1/3}{\rm V}^{-3/2}$ and $z\overline{\nu }=3/2$ and the solid one $R_{0+}\left( V_{g}\right) =R_{c}-2.4~{\rm k\Omega V}^{-1/2}\cdot\left( V_{g}-V_{gc}\right)^{1/2}$ c) $R_{0+}$ \textit{vs}. $T_{0}$ The solid line is $R_{0+}=R_{c}-37~{\rm k\Omega K}^{-1/3}\cdot T_{0}^{1/3}$.}
\label{fig9}       
\end{figure}

The full quantum scaling plot $R(T,V_{g})/R_{0\pm }\left( V_{g}\right) $ $vs.$ $\left\vert V_{g}-V_{gc}\right\vert ^{3/2}/T$ is shown in Fig.~\ref{fig10}. The upper branch stems from the NI-crossover and the lower one from the NS-transition. The two branches merge at the QSI-transition as the solid and dashed curves indicate. The solid line marks the BKT form $G_{-}\left( y\right) $ given by Eq. (\ref{eq21})) while the dashed line is $G_{+}\left( y\right) $ (Eq. (\ref{eq30})) with $f=0.065~{\rm K}^{-1/3}{\rm V}^{-3/2}$ and the $R_{0+}\left( V_{g}\right) $ estimates given in Fig.~\ref{fig9}b. The strong gate voltage dependence of $R_{0+}$ exposes clearly that two parameter scaling ($y$, $R_{0+}$) is required to enter the quantum critical regime of the variable-range hopping controlled insulator. This findings are fully analogous to those observed in the BKT-counterpart where $R_{0-}\left( V_{g}\right) $ exhibits a strong gate voltage dependence (Fig.~\ref{fig3}c) as well. The piecewise data collapse on the respective scaling function line extends roughly over two decades of the scaling argument. For large values of the scaling argument the deviations stem from the finite size effect with a limiting length which increases with reduced gate voltage (Fig.~\ref{fig3}a). In the lower branch and small scaling arguments they reveal that BKT fluctuations no longer dominate. The scaling function of the upper (Eq. (\ref{eq30})) and lower branch (Eqs. (\ref{eq17}) and (\ref{eq22})) are very different but both exhibit at the QSI-transition an essential singularity.

\begin{figure}
\includegraphics[width=\linewidth]{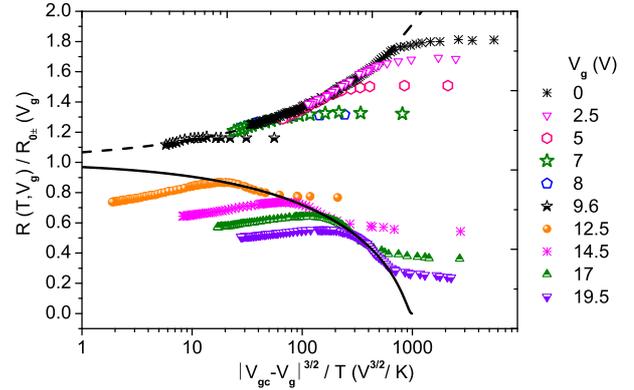}
\caption{Quantum scaling plot $R(T,V_{g})/R_{0\pm }\left( V_{g}\right) $ $vs.$ $\left\vert V_{g}-V_{gc}\right\vert ^{3/2}/T$. The lower branch taken from Fig.~\ref{fig7} originates from the NS-transition where $R_{0}(V_{g})=R_{0-}(V_{g})$. The solid line marks the BKT quantum scaling function given by Eqs. (\ref{eq13}) and (\ref{eq17}). The upper branch stems from the NI-crossover. The dashed line is the scaling form Eq. (\ref{eq29}) with $f=0.065~{\rm K}^{-1/3}{\rm V}^{-3/2}$. It is characteristic for a variable-range hopping controlled insulating ground state and determines the crossover temperature $T_{0}\left( V_{g}\right) $ (Eq. (\ref{eq30})).}
\label{fig10}       
\end{figure}

\section{Summary and Conclusions}
\label{sec:3}

Over the past twenty years the analysis of sheet resistance data taken on superconducting films, expected to undergo a thickness, magnetic field, or gate voltage tuned QSI transition, was predominantly based on the one parameter scaling form (\ref{eq18}). $z\overline{\nu }$ was usually determined by minimizing the error in data collapse using the critical sheet resistance $R_{c}$ and critical tuning parameter determined from the crossing point of the sheet resistance isotherms \cite{markovic,parendo}. Though one parameter scaling is asymptotically correct its application requires that the quantum critical regime is attained. However the generically observed saturation of the sheet resistance at low temperatures points to a finite size effect which makes this nearly impossible. Indeed our finite size scaling analysis of the BKT behavior clearly revealed that the saturation at low temperature is fully consistent with a limiting length $L$ preventing the correlation length to grow beyond $L$. Noting that disorder is irrelevant at the BKT-transition $L$ is attributable to the limited homogeneity of the films. In agreement with previous studies \cite{tsprb,tssw} we noticed that $L$ decreases substantially by approaching the QSI transition. Thus variation of the gate voltage does not affect the carrier density but the inhomogeneity landscape as well. Of course, improved homogeneity of the films would allow to circumvent these difficulties. Here we pursued a different route, we invoked two parameter with the variable $y\propto \left\vert V_{g}-V_{gc}\right\vert ^{z\overline{\nu }}/T$ replaced by $T_{c}\left( V_{g}\right) /T$ along the BKT-transition line and $T_{0}\left( V_{g}\right) /T$ along the insulator crossover line. The tuning parameter dependent amplitude $R_{0\pm }\left( V_{g}\right) $ is taken as second variable. At QSI criticality where $R_{0\pm }\left( V_{gc}\right) =R_{c}$ and $T_{c}\left( V_{g}\right) \propto T_{0}\left( V_{g}\right) \propto \left\vert V_{g}-V_{gc}\right\vert ^{z\overline{\nu }}$ the standard quantum scaling form is recovered. But given a strong gate voltage dependence of $R_{0\pm }\left( V_{g}\right) $ as encountered in the Bi film, the outlined two parameter scaling is indispensable to produce scaling plots yielding realistic estimates for $z\overline{\nu }$. Noting that $R_{0-}\left( V_{g}\right) -R_{c}$ is controlled by the amplitude of the zero
temperature correlation length $\xi \left( T=0,V_{g}\right) =\xi _{0\pm}\left( T=0\right) \left\vert V_{g}/V_{gc}-1\right\vert ^{-\overline{\nu }}$ in terms of \ $R_{0-}\left( V_{g}\right) -R_{c}\propto \xi _{0-}^{-2}\left(T=0\right) $ (Eq. (\ref{eq24})) and $R_{0+}\left( V_{g}\right) -R_{c}\propto \xi _{0+}^{-2}\left( T=0\right) $ (Eq. (\ref{eq33})), the strength of quantum fluctuations increases with reduced $\xi _{0\pm }\left( T=0\right) $, a substantial gate voltage dependence of $R_{0\pm }\left( V_{g}\right)-R_{c}$ appears to be a generic feature of the QSI-transition. According to this the two parameter scaling outlined here opens a window to attain and enter the critical regime of the QSI transition and to obtain estimates for the exponents $z\overline{\nu }$, $z$, $\overline{\nu }$ and not for $z\overline{\nu }$ only. Indeed the value of $\overline{\nu }$ is essential to clarify the relevance of disorder at zero temperature via the Harris criterion \cite{harris,aharony} and given $z$ the vlidity of hyperscaling can be verified. We are hopeful that the two parameter scaling analysis of the sheet resistance will deepen our understanding of the QSI transition in superconducting films.


\begin{thebibliography}{}

\bibitem{sondhi} Sondhi, S.L., Girvin, S.M., Carini, J.P., Shahar, D.: Continuous quantum phase transitions. Rev. Mod. Phys. \textbf{69}, 315-333 (1997)

\bibitem{tsjs} Schneider, T.,  Singer, J.M.: Phase Transition Approach to High Temperature Superconductivity. Imperial College Press, London (2000)

\bibitem{tsben} Schneider, T. in: Bennemann, K.H., Ketterson, J.B. (eds) The Physics of Superconductors, p.111. Springer, Berlin 2002)

\bibitem{markovic} Marcovi\'c, N., Christiansen, C., Mack, A., Goldman, A.M.: Superconductor-insulator transitions in 2D: The experimental situation. Phys. Status Solidi B \textbf{218}, 221-227 (2000)

\bibitem{gant} Gantmakher, V.F., Dolgopolov, V.T.: Superconductor-insulator quantum phase transition. Phys. Usp. \textbf{53}, 1-49 (2010)

\bibitem{parendo} Parendo, K.A., Tan, K.H. Sarwa B., Bhattacharya, A., Eblen-Zayas, M., Staley, N.E., Goldman, A.M.: Electrostatic tuning of the superconductor-insulator transition in two dimensions. Phys. Rev. Lett. \textbf{94}, 197004-1-4 (2005)

\bibitem{parendo2} Parendo, K.A., Tan, K.H. Sarwa B., Goldman, A.M.: Electrostatic and parallel-magnetic-field tuned two-dimensional superconductor-insulator transitions. Phys. Rev. B \textbf{73}, 174527-1-11 (2006)

\bibitem{bere} Berezinskii, V.L.: Destruction of long-range order in one dimensional and 2-dimensional systems having a continuous symmetry group: 1- classical systems. Sov. Phys. JETP \textbf{32}, 493 (1971)

\bibitem{kosterlitz} Kosterlitz, J.M., Thouless, D.J.: Ordering, metastability and phase-transitions in 2 dimensional systems. J. Phys. C \textbf{6}, 1181-1203 (1973)

\bibitem{harris} Harris, A.B.: Effect of random defects on critical behavior of ising models. J. Phys. C \textbf{7}, 1671-1692 (1974)

\bibitem{aharony} Aharony, A., Harris, A.B.: Absence of self-averaging and universal fluctuations in random systems near critical points. Phys. Rev. Lett. \textbf{77}, 3700-3703 (1996)

\bibitem{pearl} Pearl, J.: Current distribution in superconducting films carrying quantized fluxoids. Appl. Phys. Lett. \textbf{5}, 65 (1964)

\bibitem{beasley} Beasley, M.R., Mooij, J.E., Orlando, T.P.: Possibility of vortex-antivortex pair dissociation in 2-dimensional superconductors. Phys. Rev. Lett. \textbf{42}, 1165-1168 (1979)

\bibitem{nelson} Nelson, D.R., Kosterlitz, J.M.: Universal jump in superfluid density of 2-dimensional superfluids. Phys. Rev. Lett. \textbf{39}, 1201-1205 (1977)

\bibitem{privman} Privman, V. (ed): Finite-Size Scaling and Numerical Simulations of Statistical Systems. World Scientific, Singapore (1990)

\bibitem{tsprb} Schneider, T., Caviglia, A.D., Gariglio, S., Reyren, N., Triscone, J.-M.: Electrostatically-tuned superconductor-metal-insulator quantum transition at the LaAlO$_3$/SrTiO$_3$ interface. Phys. Rev. B \textbf{79}, 184502-1-9 (2009)

\bibitem{tssw} Schneider, T., Weyeneth, S.: Quantum superconductor-insulator transition: Implications of BKT-critical behavior. arXiv:1212.1330v1 (unpublished)

\bibitem{hertz} Hertz, J.A.: Quantum critical phenomena. Phys. Rev. B \textbf{14},  1165Ð1184 (1976)

\bibitem{bol} Bollinger, A.T., Dubuis, G., Yoon, J., Pavuna, D., Misewich, J., Bo\v{z}ovi\^{c}, I.: Superconductor-insulator transition in La$_{2-x}$Sr$_x$CuO$_4$ at the pair quantum resistance. Nature (London), \textbf{472}, 458-460 (2011)

\bibitem{marcano} Marcano, N., Sangiao, S., Plaza, M., P\'{e}rez, L., Pacheco, A.F., C\'{o}rdoba, R., S\'{a}nchez, M.C., Morell\'{o}n, L., Ibarra, M.R., De Teresa, J.M.: Weak-antilocalization signatures in the magnetotransport properties of individual electrodeposited Bi Nanowires. Appl. Phys. Lett., \textbf{96}, 082110-1-3 (2010)

\bibitem{bergmann} Bergmann, G.: Weak anti-localization - An experimental proof for the destructive interference of rotated spin 1/2. Solid State Commun. \textbf{42}, 815-817 (1982)

\bibitem{mott} Mott, N.F.: Conduction in non-crystalline materials. 3. Localized states in a pseudogap and near extremities of conduction and valence bands. Phil. Mag. \textbf{19}, 835 (1969)

\bibitem{fisher} Fisher, M.P.A.: Quantum phase transitions in disordered two-dimensional superconductors. Phys. Rev. Lett. \textbf{65}, 923-926 (1990)

\bibitem{weich} Weichman, P.B.: Dirty bosons: Twenty years later. Mod. Phys. Lett. B, \textbf{22}, 2623-2647 (2008)

\bibitem{fishweich} Fisher, M.P.A., Weichman, P.B., Grinstein, G., Fisher, D.S.: Boson localization and the superfluid-insulator transition. Phys. Rev. B \textbf{40}, 546-570 (1989)

\bibitem{maekawa} Maekawa, S., Fukuyama, H.: Localization effects in two-dimensional superconductors. J. Phys. Soc. Jpn. \textbf{51} 1380-1385 (1982)

\bibitem{finkel} Finkel'shtein, A.M.: Superconducting transition-temperature in amorphous films. Sov. Phys. JETP \textbf{45}, 46-49 (1987)

\bibitem{fishgrin} Fisher, M.P.A., Grinstein, G., Girvin, S.M.: Presence of quantum diffusion in two dimensions: Universal resistance at the superconductor-insulator transition. Phys. Rev. Lett. \textbf{64}, 587-590 (1990)

\bibitem{dahm} Dahm, A.J.: Further comment on "Dislocations and melting in two dimensions: The critical region". Phys. Rev. B \textbf{29}, 484-486 (1984)

\bibitem{steel} Steele, L.M., Yeager, C.J., Finotello, D.: Precision specific-heat studies of thin superfluid films. Phys. Rev. Lett. \textbf{71}, 3673-3676 (1993)

\bibitem{williams} Cho, H., Williams, G.A.: Vortex Core Size in Submonolayer Superfluid $^4$He Films. Phys. Rev. Lett. \textbf{75}, 1562-1565 (1995)

\bibitem{pierson} Pierson, S.W., Friesen, M., Ammirata, S.M., Hunnicutt, J.C., Gorham, L.A.: Dynamic scaling for two-dimensional superconductors, Josephson-junction arrays, and superfluids. Phys. Rev. B \textbf{60}, 1309-1325 (1999)

\bibitem{tstool} Schneider, T.: Estimation of the critical dynamics and thickness of superconducting films and interfaces. Phys. Rev. B \textbf{80}, 214507-1-7 (2009)

\bibitem{medv} Medvedyeva, K., Kim, B.J., Minnhagen, P.: Analysis of current-voltage characteristics of two-dimensional superconductors: Finite-size scaling behavior in the vicinity of the Kosterlitz-Thouless transition. Phys. Rev. B \textbf{62}, 14531-14540 (2000)

\bibitem{aslamosov} Aslamosov, L.G., Larkin, A.I.: Effect of fluctuations on properties of a superconductor above critical temperature. Sov. Phys. Solid State \textbf{10}, 875 (1968)

\bibitem{lee} Lee, P.A., Ramakrishnan, T.V.: Disordered electronic systems. Rev. Mod. Phys., \textbf{57} 287-337 (1985)

\bibitem{lachlan} McLachlan, D.S.: Weak-localization, spin-orbit, and electron-electron interaction effects in two- and three-dimensional bismuth films. Phys. Rev. B \textbf{28}, 6821-6832 (1983)

\bibitem{blanter} Blanter, Y.M., Vinokur, V.M., Glazman, L.I.: Weak localization in metallic granular media. Phys. Rev. B \textbf{73}, 165322-1-8 (2006)



\end{thebibliography}
\end{document}